\documentclass[epj]{svjour}

\usepackage[utf8]{inputenc}
\usepackage{amsmath}
\usepackage{url}
\usepackage{xspace}
\usepackage{graphicx}
\usepackage[bookmarksnumbered=true,bookmarksopen=true]{hyperref}
\usepackage{enumerate}

\newcommand{\Pythia}{\textsc{Pythia}}
\newcommand{\Fritiof}{\textsc{Fritiof}}

\newcommand{\dd}{\mathrm{d}}

\newcommand{\degree}{\ensuremath{^\circ}}

\newcommand{\UNIT}[1]{\ensuremath{\,{\rm #1}}\xspace}

\newcommand{\keV}{\UNIT{keV}}
\newcommand{\MeV}{\UNIT{MeV}}
\newcommand{\GeV}{\UNIT{GeV}}
\newcommand{\AGeV}{\UNIT{AGeV}}

\newcommand{\fm}{\UNIT{fm}}
\newcommand{\mb}{\UNIT{mb}}

\newcommand{\res}[3]{#1$_{#2}$(#3)}

\DeclareMathOperator{\re}{Re}

\let\oldsqrt\sqrt
\def\sqrt{\mathpalette\DHLhksqrt}
\def\DHLhksqrt#1#2{%
\setbox0=\hbox{$#1\oldsqrt{#2\,}$}\dimen0=\ht0
\advance\dimen0-0.2\ht0
\setbox2=\hbox{\vrule height\ht0 depth -\dimen0}%
{\box0\lower0.4pt\box2}}

\title{Dilepton production in proton-induced reactions at SIS energies with the GiBUU transport model}
\titlerunning{Dilepton production in proton-induced reactions}

\author{Janus Weil\inst{1}\thanks{Email: \href{mailto:janus.weil@theo.physik.uni-giessen.de}{janus.weil@theo.physik.uni-giessen.de}}, Hendrik van Hees\inst{2}, Ulrich Mosel\inst{1}}
\authorrunning{J. Weil, H. van Hees, U. Mosel}

\institute{Institut für Theoretische Physik, Universit\"at Giessen, D-35392 Giessen, Germany \and Institut für Theoretische Physik, Goethe-Universit\"at Frankfurt, D-60438 Frankfurt am Main, Germany}

\date{Received: date / Revised version: date}

\abstract{
 We present dilepton spectra from p+p, d+p and p+Nb reactions at SIS energies,
 which were simulated with the GiBUU transport model in a resonance model approach.
 These spectra are compared to the data published by the HADES and DLS collaborations.
 It is shown that the $\rho$ spectral function includes non-trivial effects already in elementary reactions, due to production via baryon resonances, which can yield large contributions to the dilepton spectrum. Dilepton spectra from nuclear reactions in the energy range of the HADES experiment are thus found to be sensitive also to properties of nucleon resonances in the nuclear medium.
\PACS{
      {PACS-key}{discribing text of that key}   \and
      {PACS-key}{discribing text of that key}
     } 
}

\begin{document}

\maketitle

\parindent0cm


\section{Introduction}

While the vacuum properties of most hadrons are known to reasonable
accuracy nowadays, it is a heavily debated question how these
properties change inside nuclear matter. In particular, various theoretical predictions
regarding the in-medium properties of the light vector mesons have been suggested.
For recent reviews on in-medium effects, see
\cite{Leupold:2009kz,Hayano:2008vn,Rapp:2009yu}.

Among these expected in-medium effects, a so-called ``collisional
broadening'' of the meson spectral function, due to
collisions with the hadronic medium, is expected.
A second class of predictions claims that the vector-meson masses
are shifted in the medium due to the partial restoration of chiral symmetry \cite{Hatsuda:1991ez}.
QCD sum rules can constrain these effects, but do not provide definitive predictions \cite{Leupold:1997dg}.

The more prominent hadronic decay modes of the vector mesons are unfavorable
for studying in-medium effects, since they are affected by
strong final-state interactions with the hadronic medium -- in
contrast to the rare dilepton decay modes. As the leptons only interact
electromagnetically, they are ideally suited to
carry the in-medium information outside to the detector,
nearly undisturbed by the hadronic medium.

Dilepton spectra from nuclear reactions with elementary projectiles have
been studied for example with the CLAS detector at JLAB, where photons with
energies of a few \GeV interact with nuclei \cite{Wood:2008ee}, or by the
E325 experiment at KEK, where 12 \GeV protons were used as projectiles
\cite{Naruki:2005kd}. On the side of the hadronic decays, most notably
$\omega\rightarrow\pi^0\gamma$ is being investigated by the CBELSA/TAPS
group in photon-induced reactions \cite{Nanova:2010sy,Nanova:2010tq}.

Early measurements of dilepton spectra from heavy-ion collisions in the low-energy regime
were conducted by the DLS collaboration \cite{Porter:1997rc}, showing an excess over the
expected yield. A similar excess was also observed in experiments at higher energies
\cite{Adamova:2002kf,Arnaldi:2006jq}, where it could be
attributed to an in-medium broadening of spectral functions \cite{vanHees:2006ng,Ruppert:2007cr,vanHees:2007th}.
For the DLS data such in-medium effects never provided a convincing explanation - a problem
that was soon known as the ``DLS puzzle'' \cite{Bratkovskaya:1997mp,Ernst:1997yy,Bratkovskaya:1998pr,Shekhter:2003xd}.

More recently, the HADES collaboration at GSI has set up an ambitious
program for measuring dilepton spectra from p+p, p+A and A+A reactions
\cite{Agakichiev:2006tg,Agakishiev:2007ts,Agakishiev:2009yf,Agakishiev:2011vf,HADES:2011ab,Agakishiev:2012tc}, in order to systematically check the old DLS data with improved statistics and to finally resolve the DLS puzzle. Up to now this endeavor has fully confirmed the validity of the DLS data and shifted the puzzle into the theory sector. It is clear that a detailed understanding of the elementary reactions is the most important prerequisite for explaining the heavy-ion data.

In this paper, we apply the Gießen Boltzmann-Uehling-Uhlenbeck
transport model (GiBUU) \cite{Buss:2011mx} to the elementary reactions
(nucleon-nucleon and proton-nucleus) studied by the HADES collaboration.
We use GiBUU to generate dilepton events and pass them through the
HADES acceptance filter, in order to compare our calculations directly
to the experimental data measured by HADES.


\section{The GiBUU transport model}

\label{sec:gibuu}

Our tool for the numerical simulation of dilepton spectra is the
GiBUU hadronic transport model, which provides a unified
framework for various types of elementary reactions on nuclei as well
as heavy-ion collisions \cite{Buss:2011mx,gibuu}. This model takes care
of the correct transport-theoretical description of the hadronic
degrees of freedom in nuclear reactions, including the propagation,
elastic and inelastic collisions and decays of particles.

In GiBUU the spectral one-particle phase-space distributions, $F(x,p)$, of all particles are obtained by solving the coupled Kadanoff-Baym equations 
\cite{KadBaym} for each particle species in their gradient-expanded form \cite{Botermans:1990qi}

\begin{equation} \label{eq:os-transp.9}
\mathcal{D} F(x,p) - \text{tr} \left\{ {\Gamma f}{\re S^{\text{ret}}(x,p)}\right\}_{\rm pb} = C(x,p)~,
\end{equation}

with

\begin{equation}
\mathcal{D} F = \left\{p_0 - H,F\right\}_{\rm pb}~.
\end{equation}

Here $\{\ldots\}_{\rm pb}$ denotes a Poisson bracket.
In the so-called backflow term (second term on the left-hand side in (\ref{eq:os-transp.9})), that is essential for off-shell transport,  $f(x,p)$ is the phase-space density related to $F$ by

\begin{equation}
F(x,p) = 2 \pi g f(x,p) \mathcal{A}(x,p) ~,
\end{equation}

where $\mathcal{A}(x,p)$ is the spectral function of the particle\footnote{$\mathcal{A}$ is normalized as $\int_0^\infty \mathcal{A}(x,p) \dd p_0 = 1$.} and $g$ is the spin-degeneracy factor. 
The quantity $\Gamma$ in the backflow term is the width of the spectral function,
and $S^{\text{ret}}(x,p)$ denotes the retarded Green's function.  Off-shell transport is thus included and leads to the correct asymptotic spectral functions of particles when they leave the nucleus. The expression $C(x,p)$ on the right-hand side of (\ref{eq:os-transp.9}) denotes the collision term that couples all particle species; it contains both a gain and a loss term. For a short derivation of this transport equation and further details we refer the reader to \cite{Buss:2011mx}. In order to solve the
BUU equation numerically, we rely on the test-particle ansatz. Here
the phase-space densities are approximated by a large number of
test particles, each represented by a $\delta$-distribution in
coordinate and momentum space.


The collision term contains all sorts of
scattering and decay processes: elastic and inelastic two-body collisions,
decays of unstable resonances and even three-body collisions. 
The two-body part of the collision term is separated into two different regimes
in terms of the available energy, $\sqrt{s}$:
a resonance model description at low energies and the \Pythia{}
string model at high energies.

For baryon-baryon collisions, the transition between the two is usually
performed at $\sqrt{s}=2.6\GeV$. There is a small window
around this border ($\pm0.2\GeV$), where both models are merged
linearly into each other in order to ensure a smooth transition.
For meson-baryon collisions, the transition region lies at
$\sqrt{s}=2.2\pm0.2\GeV$.

Unfortunately, the transition region in this default GiBUU prescription lies right inside the range of energies used for the HADES experiment. However, we think that it is important to describe all HADES spectra with one consistent model. In this paper we therefore explore the possibility of pushing the transition region up to higher energies and using an extended resonance model for all reactions measured by HADES.


In the high-energy regime the GiBUU collision term relies on the Monte Carlo event generator \Pythia{} (v6.4) \cite{pythia,Sjostrand:2006za}, which is based on the Lund string model.
Although \Pythia{} clearly has its strengths at higher energies (tens to hundreds of GeV),
it is used in GiBUU down to energies of a few \GeV. This works surprisingly
well, as has recently been demonstrated for example by GiBUU's
successful description of pion data measured by the HARP collaboration
\cite{Gallmeister:2009ht}.

Despite this good description of pion observables in the few-GeV energy regime, it turned out that the HADES dilepton data for p+p collisions at 3.5\GeV pose a somewhat greater challenge for \Pythia{} \cite{Weil:2011fa}. Most prominently, the vector-meson production is strongly overestimated by the default \Pythia{} parameters, and also the intrinsic $p_T$ distribution needs to be adjusted slightly to reproduce the HADES $p_T$ spectra.

Since a resonance description should in principle be applicable in the energy regime probed by the HADES experiment ($\sqrt{s}<3.5\GeV$), we try in the following to set up such a description as an alternative to the string model approach.


\section{The resonance-model approach}

\label{sec:RM}

The low-energy part of the nucleon-nucleon collision term is given by a resonance model based on the Teis analysis \cite{Teis:1996kx},
in which all collision cross sections are assumed to be
dominated by the excitation of baryon resonances. The GiBUU model
currently contains around 30 nucleon resonances, for a complete list see \cite{Buss:2011mx}.
However, only the subset used in the Teis analysis is actually being populated in NN collisions, see table \ref{tab:Res_par}. The properties (masses, widths and branching ratios) of all the resonances are taken from the partial-wave analysis of Manley \cite{Manley:1992yb}. All of these states, except for the $P_{33}(1600)$, are not only found in the Manley analysis, but have been confirmed, e.g., by the more recent analysis of Arndt et al.~\cite{Arndt:2006bf} and received a four-star rating from the PDG \cite{Nakamura:2010zzi}. We note already here that some of the branching ratios which are important for the present study, in particular those for decay into $\rho N$ and $\omega N$ are not very well known and still under experimental investigation \cite{Shklyar:2004ba,Anisovich:2011fc}.

\begin{table*}[t]
\begin{center}
\begin{tabular}{ | l | r c c | c c | c c c c c c c | }
  \hline
  &            & $M_0$ & $\Gamma_0$ & \multicolumn{2}{|c|}{$|\mathcal{M}^2|/16\pi$ [$\mb\GeV^2$]} & \multicolumn{7}{|c|}{branching ratio in \%} \\
  & rating     & [MeV] & [MeV]      & $NR$ & $\Delta R$ & $\pi N$ & $\eta N$ &
  $\pi\Delta$ & $\rho N$ & $\sigma N$ & $\pi N^*(1440)$ & $\sigma\Delta$\\
  \hline
  \res{P}{11}{1440} &  ****& 1462 & 391 & 70  & --- &  69 & --- &  $22_P$     & ---        &  9  & --- & --- \\
  \res{S}{11}{1535} &   ***& 1534 & 151 &  8  & 60  &  51 &  43 & ---         & $2_S+1_D$  &  1  &  2  & --- \\
  \res{S}{11}{1650} &  ****& 1659 & 173 &  4  & 12  &  89 &  3  &  $2_D$      & $3_D$      &  2  &  1  & --- \\
  \res{D}{13}{1520} &  ****& 1524 & 124 &  4  & 12  &  59 & --- &  $5_S+15_D$ & $21_S$     & --- & --- & --- \\
  \res{D}{15}{1675} &  ****& 1676 & 159 & 17  & --- &  47 & --- &  $53_D$     & ---        & --- & --- & --- \\
  \res{P}{13}{1720} &     *& 1717 & 383 &  4  & 12  &  13 & --- & ---         & $87_P$     & --- & --- & --- \\
  \res{F}{15}{1680} &  ****& 1684 & 139 &  4  & 12  &  70 & --- &  $10_P+1_F$ & $5_P+2_F$  &  12 & --- & --- \\
  \hline
  \res{P}{33}{1232} &  ****& 1232 & 118 & OBE & 210 & 100 & --- & ---         & ---        & --- & --- & --- \\
  \res{S}{31}{1620} &    **& 1672 & 154 &  7  &  21 &   9 & --- &  $62_D$     & $25_S+4_D$ & --- & --- & --- \\
  \res{D}{33}{1700} &     *& 1762 & 599 &  7  &  21 &  14 & --- &  $74_S+4_D$ & $8_S$      & --- & --- & --- \\
  \res{P}{31}{1910} &  ****& 1882 & 239 & 14  & --- &  23 & --- & ---         & ---        & --- &  67 & $10_P$ \\
  \res{P}{33}{1600} &   ***& 1706 & 430 & 14   & --- &  12 & --- &  $68_P$     & ---        & --- &  20 & --- \\
  \res{F}{35}{1905} &   ***& 1881 & 327 &  7  &  21 &  12 & --- &  $1_P$      & $87_P$     & --- & --- & --- \\
  \res{F}{37}{1950} &  ****& 1945 & 300 & 14  & --- &  38 & --- &  $18_F$     & ---        & --- & --- & $44_F$ \\
  \hline
\end{tabular}
\end{center}
\caption{Resonance parameters according to Manley \cite{Manley:1992yb} (columns 2-4), together with matrix elements for production in pp collisions (columns 5 and 6) and branching ratios of the resonance decay modes (columns 7-13). Subscripts indicate the relative angular momentum of the outgoing particles in the respective decay channel.}
\label{tab:Res_par}
\end{table*}

We use all the resonance parameters and branching ratios exactly as given by Manley, with one exception: The $\rho\Delta$ decay channels are introduced by Manley only in order to account for missing inelasticities, which are not covered by one- and two-pion final states. In that sense, Manley has no real evidence for the $\rho\Delta$ final state in particular, but just uses this decay channel to account for the left-over strength. Therefore we take the freedom to replace the $\rho\Delta$ decays by $\sigma\Delta$, in order to avoid an overestimation of the $\rho$-meson production. The influence of Manley's $\rho\Delta$ decay channels on dilepton spectra was already discussed in \cite{Effenberger:1999nn} for the case of pion-induced reactions. The dilepton spectra actually give a hint that the needed $3\pi$ inelasticity might not be in the $\rho\Delta$, but instead in some other channel, as e.g. $\sigma\Delta$.

Also the width parametrizations are taken from the Manley analysis, where the partial widths for, e.g., $\Delta\rightarrow\pi N$ and $\rho\rightarrow\pi\pi$ are parametrized according to

\begin{equation} \label{eq:Gamma_pipi}
 \Gamma(m) = \Gamma_0 \frac{m_0}{m} \left(\frac{q}{q_0}\right)^3 \frac{q_0^2+\Lambda^2}{q^2+\Lambda^2} .
\end{equation}

Here $m_0$ is the mother particle's pole mass, $m$ is its off-shell mass, $\Gamma_0$ is the on-shell width (at $m=m_0$); $q$ denotes the final-state center-of-mass momentum for mass $m$, while $q_0$ is the same quantitiy for mass $m_0$, and $\Lambda=1/R=1\fm^{-1}$ can be viewed as a cutoff-parameter. It has been shown in \cite{effe_phd}, that eq.~(\ref{eq:Gamma_pipi}) gives a good description of the experimental phase shifts in $\pi\pi$ and $\pi N$ scattering. For the detailed treatment of the other decay channels, we refer to chapter 3.3.1 of \cite{Buss:2011mx}.

The resonance model used in this work is based on the Teis model, but modifies and extends it in several aspects. We take into account the following nucleon-nucleon scattering channels:

\begin{enumerate}
 \item $NN\rightarrow NN$
 \item $NN\rightarrow N\Delta$,
 \item $NN\rightarrow NN^*,\; N\Delta^*$,
 \item $NN\rightarrow \Delta\Delta$,
 \item $NN\rightarrow \Delta N^*,\; \Delta\Delta^*$,
 \item $NN\rightarrow NN\pi$ (non-res. BG)
 \item $NN\rightarrow NN\omega,\; NN\pi\omega,\; NN\phi$ (non-res.),
 \item $NN\rightarrow BYK$ (with $B=N,\Delta$; $Y=\Lambda,\Sigma$).
\end{enumerate}

For the elastic cross sections (first item), we rely on the parametrizations by Cugnon et al.~\cite{Cugnon:1996kh} (for beam momenta below $p_{\rm lab}\approx2.776\GeV$) and the PDG \cite{Montanet:1994xu} (above). For details see also \cite{Buss:2011mx}.

The single-resonance excitation channels (items 2 and 3) were already included in the Teis analysis.
While the $N\Delta$ channel is treated by an OBE model according to Dmitriev et al. \cite{Dmitriev:1986st}, the higher resonances are produced in a pure phase-space approach with constant matrix elements,

\begin{equation}
 \sigma_{NN\rightarrow NR}=\frac{C_I}{p_is}\frac{|\mathcal{M}_{NR}|^2}{16\pi}\int\dd\mu \mathcal{A}_R(\mu)p_F(\mu) .
\end{equation}

Here, $p_i$ and $p_F$ denote the center-of-mass momenta in the initial and final state, respectively.
The matrix elements, $\mathcal{M}_{NR}$, have previously been fitted by Teis to exclusive meson production ($\pi$, $2\pi$, $\eta$ and $\rho$). Our values are listed in tab. \ref{tab:Res_par}. $\mathcal{A}_R$ denotes the resonance spectral function,

\begin{equation}
 \mathcal{A}_R (\mu) = \frac{2}{\pi} \frac{\mu^2\Gamma_R(\mu)}{(\mu^2-M_R^2)^2+\mu^2\Gamma_R^2(\mu)} \; .
\end{equation}

In principle all production channels are assumed to be isospin-symmetric, with the Clebsch-Gordan factors, $C_I$, resulting from this symmetry. The only exception from this isospin symmetry is the $S_{11}(1535)$ resonance: The exclusive $\eta$ production, which is assumed to proceed exclusively via this resonance, is known to be significantly larger for $pn$ than for $pp$ \cite{Calen:1998vh}, therefore we use

\begin{equation}
 |\mathcal{M}_{pn\rightarrow NN^*(1535)}|^2 = 6.5 \cdot |\mathcal{M}_{pp\rightarrow NN^*(1535)}|^2.
\end{equation}

Note that while the $S_{11}(1535)$ is known to dominate the $\eta$ production in pp at low energies, there may of course be other contributions \cite{Balestra:2004kg}.

The single-pion production cross section can not be described satisfactorily by resonance contributions alone, and one has to add a non-resonant background term \cite{Buss:2011mx,Teis:1996kx} (slightly refitted here), whose largest contributions appear on the left-hand shoulder of the $N\Delta$ peak.

\begin{figure*}[t]
  \includegraphics[width=\textwidth]{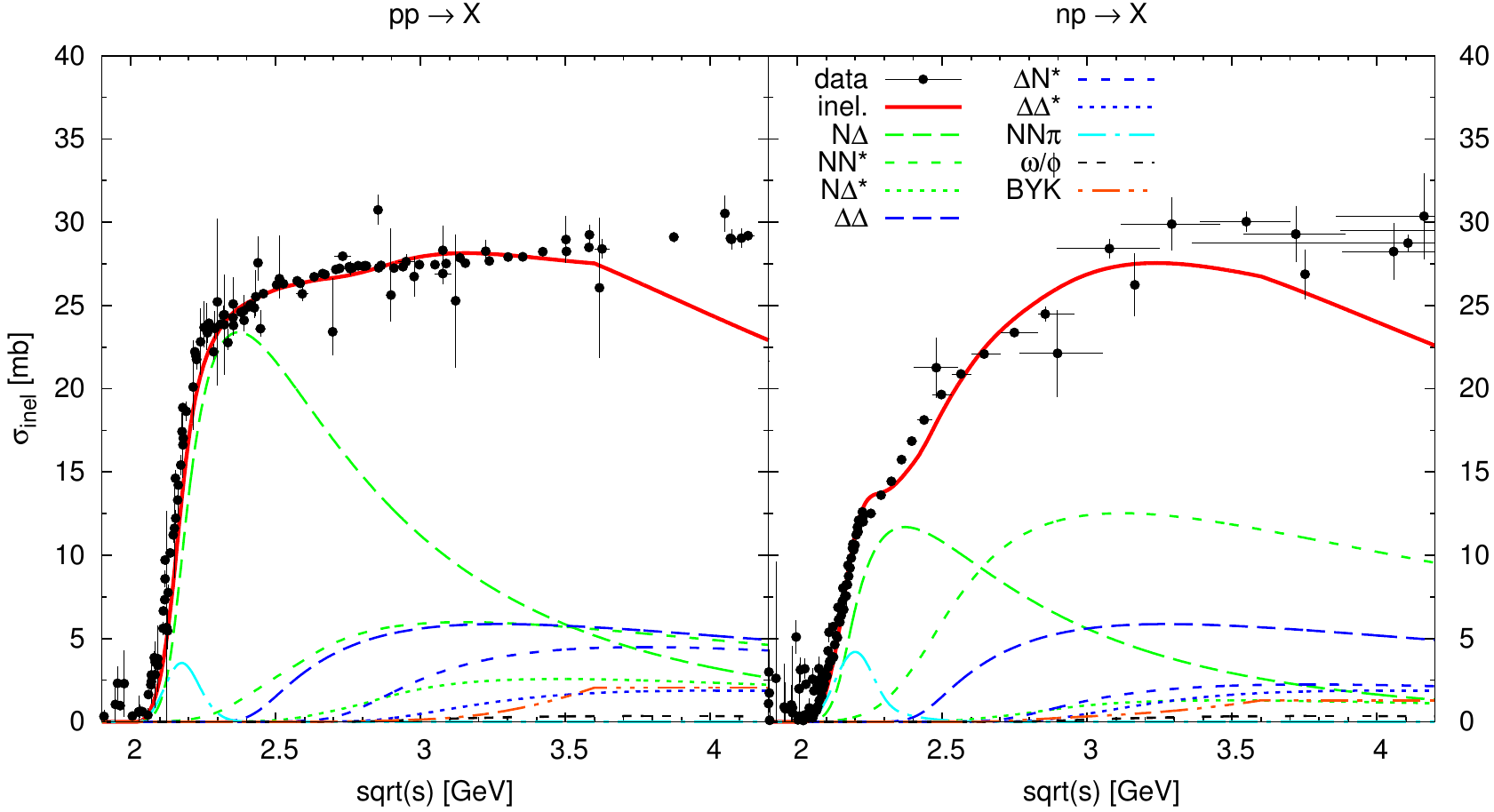}
  \caption{(Color online) Inelastic pp and pn cross sections in the resonance model. The data points shown here have been obtained by subtracting the parametrized elastic cross section from the total cross section data \cite{Nakamura:2010zzi}.}
  \label{fig:NN_inel}
\end{figure*}

Most of the resonance-production matrix elements are adopted from Teis. However, me make a few modifications. In particular we reduce the contributions of the $D_{15}(1675)$, $P_{31}(1910)$ and $P_{33}(1600)$, which were extremely large in the Teis analysis, in favor of the $P_{11}(1440)$ and double-$\Delta$ contributions. This gives an improved threshold behavior of the $2\pi$ production channels (in line with the analysis of  Cao et al.~\cite{Cao:2010km}), as well as a better agreement with the inelastic nucleon-nucleon cross sections, cf.~fig.~\ref{fig:NN_inel}.

Furthermore, we add another isospin-asymmetry factor for the $P_{11}(1440)$ state:

\begin{equation}
 |\mathcal{M}_{pn\rightarrow NN^*(1440)}|^2 = 2 \cdot |\mathcal{M}_{pp\rightarrow NN^*(1440)}|^2.
\end{equation}

This is done in order to improve the agreement with the np inelastic cross section data, which would otherwise be underestimated significantly.

Also the double-resonance production (items 4 and 5), which in the Teis model was limited to $\Delta\Delta$, is performed in a phase-space approach, analogous to the single-resonance excitation:

\begin{align}
 \sigma_{NN\rightarrow \Delta R} & = \frac{C_I}{p_is}\frac{|\mathcal{M}_{\Delta R}|^2}{16\pi} \notag\\
 & \times \int\dd\mu_1\dd\mu_2 \mathcal{A}_{\Delta}(\mu_1)\mathcal{A}_{R}(\mu_2)p_F(\mu_1,\mu_2).
\end{align}

Here one integrates over the spectral functions of both resonances ($\mu_{1,2}$ being their masses).

In the Teis analysis, the production mechanisms were restricted to $NN\rightarrow NR$ and $NN\rightarrow\Delta\Delta$, so the obvious extension candidate would be general double-resonance excitation channels ($NN\rightarrow R_1 R_2$). The channels taken into account by Teis were fitted to single- and double-pion production data. Therefore his model is only guaranteed to work in the low-energy region. At higher energies, the model starts to fail, since the more inclusive multi-meson final states are not included.
If we want to describe $NN$ collisions in the HADES energy regime of $\sqrt{s}\approx2.4-3.2\GeV$ with a resonance model, we clearly need to extend the Teis approach. 
We do this by restricting ourselves to the same set of resonances (cf. tab.~\ref{tab:Res_par}), but extending the production mechanisms.

Since Teis already describes the exclusive $\pi$, $2\pi$, $\rho$ and $\eta$ production, what is missing are channels like e.g. $\pi\eta$, $\pi\rho$, $3\pi$, $2\eta$, $2\rho$, etc. Unfortunately there are almost no experimental data available for these channels. We thus have to rely on the cross sections obtained from \Pythia{} as an estimate to fix these channels. According to \Pythia{}, the inclusive $\rho$ and $\eta$ production is in fact dominated by the channels $\pi\rho$ and $\pi\eta$, respectively, at the highest HADES energy of $\sqrt{s}\approx3.2\GeV$. Therefore we concentrate on these two for now, and neglect all others. Our strategy to satisfy these channels relies on double-resonance excitation, $NN\rightarrow \Delta R$, where the $\Delta$ decays into $\pi N$, while the other resonance $R$ will be one with an $\eta N$ or $\rho N$ decay channel, so that we end up with a $\pi\eta$ or $\pi\rho$ final state (note that we do not include cascade decays of single resonances into $\pi\eta N$, as treated 
for example in \cite{Kashevarov:2009ww}, since our model misses the corresponding decays modes, such as $\eta\Delta$). We add three new classes of production channels:

\begin{enumerate}[i)]
 \item $NN\rightarrow\Delta S_{11}(1535)$ ($\rightarrow NN\pi\eta$),
 \item $NN\rightarrow\Delta N^*$ ($\rightarrow NN\pi\rho$), \\
       $N^*=D_{13}(1520),S_{11}(1650),F_{15}(1680),P_{13}(1720)$,
 \item $NN\rightarrow\Delta\Delta^*$ ($\rightarrow NN\pi\rho$), \\
       $\Delta^*=S_{31}(1620),D_{33}(1700),F_{35}(1905)$.
\end{enumerate}

For each of these we need one new parameter, namely the matrix elements, $|\mathcal{M}_1|^2/16\pi=60\mb\GeV^2$, $|\mathcal{M}_2|^2/16\pi=12\mb\GeV^2$ and $|\mathcal{M}_3|^2/16\pi=21\mb\GeV^2$, as listed in tab.~\ref{tab:Res_par}. As noted before, we fix the matrix elements to roughly fit the \Pythia{} cross sections for $\pi\eta$ and $\pi\rho$ production (with further constraints from the total $pp$ cross section as well as the HADES dilepton data). As for the exclusive production, we assume that the $\eta$ meson is produced exclusively via the $S_{11}(1535)$, while the $\rho$ production proceeds via a number of $N^*$ and $\Delta^*$ resonances. It should be noted that the new channels will not affect the exclusive meson production fitted by Teis, except for the $2\pi$ channel, which gets minor contributions from these channels.

The production of $\omega$ and $\phi$ mesons is not carried out via baryonic resonances in our model (although a coupling of the $\omega$ to nucleon resonances has been reported for example in \cite{Mosel:2001fr,Post:2001am,Penner:2002md,Shklyar:2004ba}). Instead, their production cross sections are parametrized in a phenomenological manner \cite{Sibirtsev:1996uy}:

\begin{equation}
 \sigma(pp\rightarrow ppV) = a (1-x)^b x^c, \;\;\; \text{with} \; x=s_0/s.
\end{equation}

Here $s_0=(2m_N+m_V)^2$ is the threshold energy, and the parameters $a$, $b$ and $c$ are listed in table \ref{tab:NNV}. We use this parametrization not only for exclusive $\omega$ and $\phi$ production, but also for $NN\rightarrow NN\pi\omega$. Since there are no data available for this channel, we fitted its parameters to the \Pythia{} results.

\begin{table}[h]
  \begin{center}
    \begin{tabular}{|c|c|c|c|c|c|}
      \hline
                  & $\sqrt{s_0}$ [GeV] & $a$ [mb] & $b$  & $c$ & Ref.  \\
      \hline
      $\omega$    & 2.658              & 5.3      & 2.3  & 2.4 & \cite{AbdelBary:2007sq} \\
      $\pi\omega$ & 2.796              & 1.0      & 1.5  & 1.1 & - \\
      $\phi$      & 2.895              & 0.01     & 1.26 & 1.66& \cite{Paryev:2008ck} \\
      \hline
    \end{tabular}
  \end{center}
  \caption{Parameters for vector-meson production.}
  \label{tab:NNV}
\end{table}

As seen in fig.~\ref{fig:NN_inel}, we achieve a good agreement with data for the inelastic pp cross section up to about $\sqrt{s}=3.5\GeV$. At higher energies $3\pi$ and $4\pi$ production becomes important, which is underestimated by our model (and other channels which we miss completely). In the np cross section there are minor deviations, and unfortunately also the quality of the data is not quite as good as for pp.


\section{Dilepton decays and form factors}

In the GiBUU model the following dilepton decay modes are taken into account:

\begin{itemize}
\item direct decays, as $V \rightarrow e^+e^-$, \\
  with $V=\rho^0,\omega,\phi$\quad or\quad $\eta \rightarrow e^+e^-$\;,
\item Dalitz decays, as $P \rightarrow e^+e^-\gamma$ with
  $P=\pi^0,\eta$\quad \\ or\quad $\omega \rightarrow \pi^0e^+e^-$\quad or
  \quad $\Delta \rightarrow Ne^+e^-$\;.
\end{itemize}

Most of them are
treated similarly as in \cite{effe_phd}. The leptonic decay widths of
the vector mesons are taken under the assumption of strict
vector-meson dominance (VMD),

\begin{equation} \label{eq:gamma_ee}
  \Gamma_{V\rightarrow e^+e^-}(\mu)=C_V\frac{m_V^4}{\mu^3},
\end{equation}

where $\mu$ is the meson's off-shell mass, $m_V$ is the pole mass, and the constants $C_V$ are listed in table \ref{tab:vm_dil} (taken from \cite{Nakamura:2010zzi}).
Although the physical threshold of the dileptonic decay channels of course lies
at $2m_e$, contributions of $\rho$ mesons below $m=2m_\pi$ are frequently neglected in transport simulations.
The reason for this artificial threshold is purely numerical: The $\rho$ spectral function has a sharp drop at the $2\pi$ threshold, and it is numerically very difficult to populate the spectral function below this threshold, where it is almost vanishing. Here we make additional numerical efforts to include the contribution of $\rho$ mesons below the $2\pi$ threshold, since it can give significant contributions to the total dilepton spectrum for certain reactions.

\begin{table}[h]
  \begin{center}
    \begin{tabular}{|c|c|c|c|c|c|}
      \hline
      $V$ & $m_V (\MeV)$ & $\Gamma_{ee} (\keV)$ & $C_V=\Gamma_{ee}/m_V$ \\
      \hline
      $\rho$   & 775.49   & 7.04 & $9.078\cdot10^{-6}$ \\
      $\omega$ & 782.65   & 0.60 & $7.666\cdot10^{-7}$ \\
      $\phi$   & 1019.455 & 1.27 & $1.246\cdot10^{-6}$\\
      \hline
    \end{tabular}
  \end{center}
  \caption{Dilepton-decay constants for $V\rightarrow e^+e^-$.}
  \label{tab:vm_dil}
\end{table}

While the direct decay of the $\eta$ meson into a $\mu^+\mu^-$ pair
has been observed, for the corresponding $e^+e^-$ decay only an upper
limit of $\mbox{BR}(\eta\rightarrow e^+e^-)<2.7\cdot10^{-5}$ is known
\cite{Berlowski:2008zz}. In fact this limit has been pushed down to $4.9\cdot10^{-6}$ lately using HADES dilepton data \cite{HADES:2011ab}. However, the theoretical expectation from
helicity suppression is still four orders of magnitude lower
\cite{Browder:1997eu}. The absence of any $\eta$ peak in the measured spectra allows us to conclude that
the true branching ratio must be significantly lower than the upper limit just mentioned \cite{muehlich_phd}.
Therefore we do not include the $\eta\rightarrow e^+e^-$ decay in our analysis.

The Dalitz decays of the pseudoscalar mesons, $P=\pi^0,\eta$,
are treated via the parametrization \cite{Landsberg:1986fd},
\begin{align}
  \frac{\dd\Gamma_{P\rightarrow\gamma e^+e^-}}{\dd\mu} = &
  \frac{4\alpha}{3\pi}\frac{\Gamma_{P\rightarrow\gamma\gamma}}{\mu}
  \left(1-\frac{\mu^2}{m_P^2}\right)^3 |F_P(\mu)|^2,
\end{align}
with $\Gamma_{\pi^0\rightarrow\gamma\gamma}=7.8\cdot10^{-6}\MeV$,
$\Gamma_{\eta\rightarrow\gamma\gamma}=4.6\cdot10^{-4}\MeV$ and the
form factors,
\begin{alignat}{4}
  F_{\pi^0}(\mu) & = 1 + b_{\pi^0}\mu^2, \quad & b_{\pi^0} & = 5.5\GeV^{-2} \;, \\
  F_{\eta}(\mu) & = \left(1-\frac{\mu^2}{\Lambda_\eta^2}\right)^{-1},
  \quad & \Lambda_\eta & = 0.676\GeV \;.
\end{alignat}

The above value of $\Lambda_\eta$ has been recently determined
from the HADES data at 2.2 \GeV beam energy \cite{spruck_phd} and agrees reasonably well
with the values found by NA60 \cite{:2009wb} and CB/TAPS \cite{Berghauser:2011zz}.
As shown in \cite{Bratkovskaya:1995kh}, the dilepton decays of the pseudoscalar mesons
is expected to follow an anisotropic angular distribution,

\begin{align}
 \frac{\dd\Gamma_{P\rightarrow\gamma e^+e^-}}{\dd\cos\theta}\propto1+\cos^2(\theta),
\end{align}

where $\theta$ is the angle of the electron momentum with respect to the dilepton momentum.
This has been confirmed recently by HADES data \cite{HADES:2012aa}.
All other decays are treated isotropically in our model.

The parametrization of the $\omega$ Dalitz decay,

\begin{align}
  \frac{\dd\Gamma_{\omega\rightarrow\pi^0e^+e^-}}{\dd\mu} & = \frac{2\alpha}{3\pi}\frac{\Gamma_{\omega\rightarrow\pi^0\gamma}}{\mu} \notag \\
  & \times \left[ \left(1+\frac{\mu^2}{\mu_\omega^2-m_\pi^2}\right)^2 -\frac{4\mu_\omega^2\mu^2}{(\mu_\omega^2-m_\pi^2)^2} \right]^{3/2} \notag \\
  & \times |F_\omega(\mu)|^2, \\
  |F_\omega(\mu)|^2 & = \frac{\Lambda_\omega^4}{(\Lambda_\omega^2-\mu^2)^2+\Lambda_\omega^2\Gamma_\omega^2},
\end{align}

is adopted from \cite{Bratkovskaya:1996qe,effe_phd} with
$\Gamma_{\omega\rightarrow\pi^0\gamma}=0.703\MeV$,
$\Lambda_\omega=0.65\GeV$ and $\Gamma_\omega=75\MeV$. Here we note
that the form factor of the $\omega$ Dalitz decay is also
well-constrained by data \cite{:2009wb}.

For the $\Delta$-Dalitz decay, we use the parametrization from \cite{Krivoruchenko:2001hs},
\begin{align}
  \frac{\dd\Gamma_{\Delta\rightarrow Ne^+e^-}}{\dd\mu} & = \frac{2\alpha}{3\pi\mu}\Gamma_{\Delta\rightarrow N\gamma^*}, \\
  \Gamma_{\Delta\rightarrow N\gamma^*} & =
  \frac{\alpha}{16}\frac{(m_\Delta+m_N)^2}{m_\Delta^3m_N^2}
  \left[(m_\Delta+m_N)^2-\mu^2\right]^{1/2} \notag \\
  & \times \left[(m_\Delta-m_N)^2-\mu^2\right]^{3/2} |F_\Delta(\mu)|^2,
\end{align}

where we neglect the electron mass. The electromagnetic N-$\Delta$ transition
form factor $F_\Delta(\mu)$ is an issue of
ongoing debate. Unlike the other semileptonic Dalitz decays, it is
poorly constrained by data. At least at the real-photon point
($\mu=0$) it is fixed by the decay width
$\Gamma_{\Delta\rightarrow N\gamma}\approx0.66\MeV$ \cite{Nakamura:2010zzi}
to $|F_\Delta(0)|=3.03$, and also in the space-like region this form
factor is well-constrained by electron-scattering data on the
nucleon. However, it is basically unknown in the time-like regime,
which is being probed by the $\Delta$ Dalitz decay.

Theoretical models for the N-$\Delta$ transition form factor usually assume one or more VMD-inspired peaks in the time-like region \cite{Krivoruchenko:2001jk,Caia:2004pm,Wan:2005ds,Ramalho:2012ng}. However, the data in the space-like region does not provide sufficient constraints to fix the behavior in the time-like region.

Moreover, a VMD-like $\Delta$ form factor would imply a coupling of the $\Delta$ to the $\rho$ meson, which has never been observed directly and could only play a role far off the $\Delta$ pole, where its strength is completely unknown \cite{Post:2000qi}.

In order to demonstrate the uncertainty connected to this form factor, we will in the following use as an example the model of \cite{Wan:2005ds}. However, we note that recently a new form-factor calculation has appeared \cite{Ramalho:2012ng}, whose results differ significantly from the ones given in \cite{Wan:2005ds}.

For the other baryonic resonances we don't explicitly include a Dalitz decay, but evaluate their contributions to the dilepton spectrum through the two-step process $R\rightarrow N\rho\rightarrow Ne^+e^-$. In the transport-typical manner we cut the corresponding diagrams, separating the production and decay vertices of the resonance and neglecting any phases and interferences.
Below the $2\pi$ threshold, the $\rho$ meson width becomes very small because here only the electromagnetic decay width is active. This smallness of the width, however, is counteracted to some degree by the propagator of the virtual photon that enhances small dilepton masses, see eq.~(\ref{eq:gamma_ee}). In an alternative treatment, in which the $N^*$ resonances undergo direct Dalitz decay, an electromagnetic form factor at the $N N^* \gamma^*$ vertex would mimick the $\rho$ propagator. These two methods are fully equivalent if the phase relations between the decaying resonance and the dileptons can be neglected and a corresponding form factor is used (our model relies on the assumption of strict VMD). Any interaction of the $\rho$ meson between its production and decay, leading to a broadening of the $\rho$ spectral function, could be absorbed into a medium dependence of the form factor.

Further we include pn-Bremsstrahlung in phase-space corrected soft-photon approximation \cite{Gale:1988yk,Wolf:1990ur}, which can be written as

\begin{align}
 \frac{\dd\sigma_{pn\rightarrow pne^+e^-}}{\dd M\dd E\dd\Omega} & = \frac{\alpha^2}{6\pi^3} \frac{q}{ME^2} \bar{\sigma}(s) \frac{R_2(s_2)}{R_2(s)} \;, \\
 \bar{\sigma}(s) & = \frac{s-(m_1+m_2)^2}{2m_1^2}\sigma^{pn}_{el}(s) \;, \\
 R_2(s) & = \sqrt{1-(m_1+m_2)^2/s} \;, \\
 s_2 & = s + M^2 - 2E\sqrt{s} \;,
\end{align}

where $M$ is the mass of the dilepton pair, $q$, $E$ and $\Omega$ are its momentum, energy and solid angle in the pn center-of-mass frame and $s$ is the Mandelstam's variable. Further, $m_1$ is the mass of the charged particle (proton), $m_2$ is the mass of the neutral particle (neutron) and $\sigma^{pn}_{el}$ is the elastic pn cross section.

pp-Bremsstrahlung can not be treated in this simple approximation, since it involves a destructive interference between the graphs involved. Due to this interference it is much smaller than the pn-Bremsstrahlung and therefore is being neglected here.

Further we note that also the Bremsstrahlung contribution involves a form factor, i.e. the time-like nucleon form factor. Just as the $\Delta$ transition form factor, it is not well-constrained in the time-like region and is usually neglected, also in recent OBE models \cite{Shyam:2003cn,Kaptari:2005qz,Shyam:2010vr}. 


\section{Dilepton spectra from elementary N+N collisions}

After describing the basic ingredients of the model, we will now present simulated dilepton spectra for various  elementary reactions (p+p and d+p). The HADES collaboration has performed measurements of dilepton
spectra from elementary p+p reactions at the beam energies listed in table \ref{tab:HADES_reactions}.

\begin{table}[h]
  \begin{center}
    \begin{tabular}{|c|c|c||c|c|}
      \hline
      $E_{\rm kin}$ & $\sqrt{s}$ & $p_{\rm lab}$ & $p_{\rm lep}^{\rm min}$ & $p_{\rm lep}^{\rm max}$ \\
      \hline
      1.25 & 2.42 & 1.98 & 0.05 & 1.8 \\
      2.20 & 2.76 & 2.99 & 0.10 & 2.0 \\
      3.50 & 3.17 & 4.34 & 0.08 & 2.0 \\
      \hline
    \end{tabular}
  \end{center}
  \caption{Kinematic conditions of the elementary collisions measured by HADES and corresponding cuts on the single lepton momenta (all numbers in GeV).}
  \label{tab:HADES_reactions}
\end{table}

For the beam energy of 1.25\AGeV, also dp collisions have been measured.
All simulated spectra have been filtered with the HADES acceptance filter (HAFT, version 2.0) \cite{hades,galatyuk_priv}, in order to take care of the geometrical acceptance and resolution of the detector. In addition, a dilepton opening angle cut of $\theta_{ee}>9\degree$ is applied in all cases, as well as the single-lepton momentum cuts listed in table \ref{tab:HADES_reactions}, matching the experimental analysis procedure.


\subsection{p + p at 1.25 GeV}

\begin{figure}[ht]
  \begin{center}
  \includegraphics[width=0.5\textwidth]{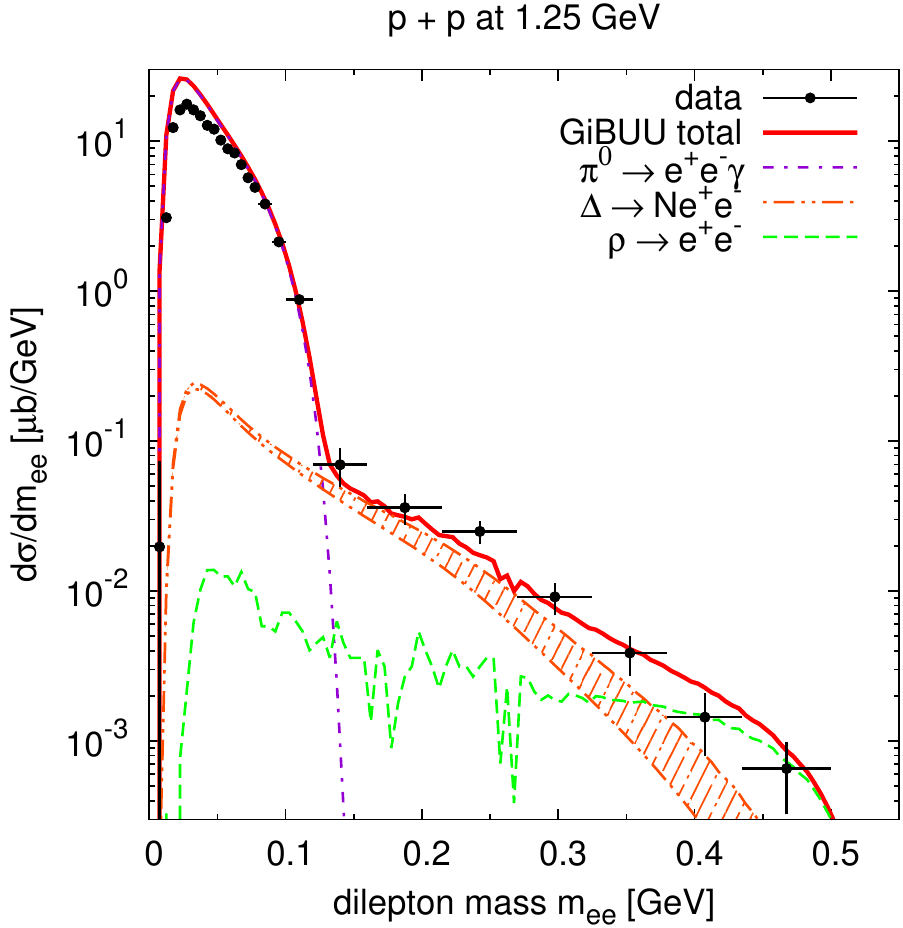}
  \end{center}
  \caption{(Color online) Dilepton mass spectrum for pp at 1.25\GeV, in comparison to the data from \cite{Agakishiev:2009yf}. The different contributions are indicated in the figure. The hatched area indicates the effect of the $\Delta$ form factor.}
  \label{fig:pp125}
\end{figure}

The lowest HADES energy, $E_{\rm kin}=1.25\GeV$, corresponding to $\sqrt{s}\approx2.4\GeV$, is just below the $\eta$ production threshold, and also for $\rho$ mesons there is only a small sub-threshold contribution from the low-mass tail of the $\rho$ spectral function.

This means that the dilepton spectrum is dominated by the $\pi^0$ and $\Delta$ Dalitz decays. One should note that at this energy, almost all pions are produced via excitation and decay of the $\Delta$ resonance.

Both of these Dalitz decays involve a transition form factor. But while the form factor of the $\pi^0$ Dalitz channel has been determined experimentally to a reasonable precision \cite{Landsberg:1986fd}, the electromagnetic transition form factor of the $\Delta$ Dalitz decay is basically unknown in the time-like region (cf. previous sect.).

However, the dilepton spectrum at $E_{\rm kin}=1.25\GeV$ is only mildly sensitive to this form factor, since the energy is not large enough to reach the VM pole-mass region. As fig. \ref{fig:pp125} shows, the simulation profits from including a form factor (shaded band) which exhibits a moderate rise in the time-like region of small $q^2$, but it is not sensitive to the actual VMD peak of such a form factor. Here we have used the form factor from \cite{Wan:2005ds}, but we have also verified that using a standard VMD form factor yields virtually the same results for this energy.

It is interesting to note that other calculations achieve a good agreement with the HADES data for pp collisions at 1.25\GeV without including any form factor for the $\Delta$ Dalitz channel \cite{Shyam:2010vr} (which might be partly due to the different width parametrization used).

Furthermore we note that the slight overshooting in the pion channel is apparently due to the higher resonances, which are produced only in phase-space approximation. The missing treatment of proper angular momentum distributions seems to interfere with the HADES acceptance here, however it does not seem to be a problem at higher energies. We have verified that the discrepancy disappears if all pions are produced exclusively via $\Delta$ excitation, neglecting contributions from higher resonances.


\subsection{d + p at 1.25 GeV}

\begin{figure}[ht]
  \begin{center}
  \includegraphics[width=0.5\textwidth]{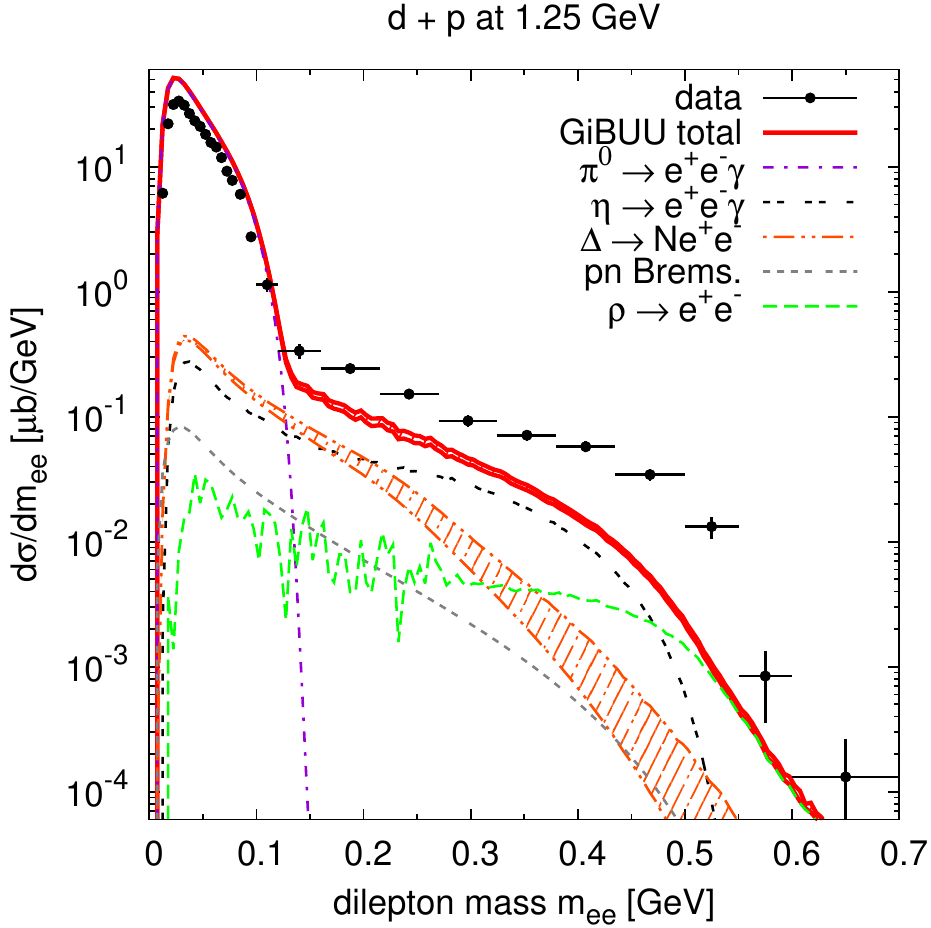}
  \end{center}
  \caption{(Color online) Dilepton mass spectrum for d+p at 1.25\GeV, in comparison to the data from \cite{Agakishiev:2009yf}.}
  \label{fig:dp125}
\end{figure}

In addition to the proton beam, also a deuteron beam with a kinetic energy of 1.25\AGeV has been used by HADES. Here, a trigger on forward-going protons has been set up in order to select the (quasi-free) np collisions, which are only accessible in this way.

Due to the motion of the bound nucleons in the deuteron, the energy of the NN collisions is smeared out here, compared to the proton-beam case, with a tail reaching above the $\eta$-production threshold. The momentum distribution of the nucleons is determined by the deuteron potential, which in our simulations is given by the Argonne V18 potential \cite{Wiringa:1994wb}.

Fig.~\ref{fig:dp125} shows the dilepton invariant mass spectrum for this reaction. While the $\pi^0$-Dalitz channel in the low-mass region shows a similarly good agreement as in the pp case, the data points at larger invariant masses are underestimated by a factor of two or more.

A stronger $\Delta$ channel can apparently not explain the shoulder in the data around 500\MeV, since it falls off too steeply, even when including a form factor. In addition to the enhanced $\eta$ production in $\rm np\rightarrow np\eta$, as described earlier, we have included a $\rm np\rightarrow d\eta$ channel, which dominates the $\eta$ production from np at threshold \cite{Calen:1998vh}.

Unfortunately, the strong $\rm pn\rightarrow pn\rho^0$ channel is experimentally not so well known. In our model, the $\rho^0$ production in d+p at 1.25\GeV is dominated by the $D_{13}(1520)$ and $S_{11}(1535)$ resonances. The latter is enhanced in np (because of its dominant role in $\eta$ production). The former is assumed to be isospin-symmetric, which may not be the case.

In an OBE-model study \cite{Shyam:2010vr} it has been found that the radiation from internal pion lines (with the appropriate VMD form factor) gives a sizable contribution at large invariant masses. Such a diagram implicitly contains a $\rho^0$ propagator (through the form factor), and gives additional $\rho$-like contributions on top of the resonance contributions included in our model.

Moreover, we might underestimate the `pure' Bremsstrahlung contributions, which do not involve resonance excitations, due to the soft-photon approximation. However, it is not expected that these terms would yield any dominant contributions \cite{Schafer:1994tz,deJong:1996jm,Shyam:2010vr}.

As recently argued in \cite{Martemyanov:2011hv}, the inclusion of a ``radiative capture'' channel $\rm np\rightarrow de^+e^-$, fixed via deuteron photo-disintegration, might give further contributions in the high-mass region.

According to our analysis, the most probable candidate to fill the missing yield are indeed $\rho$-like contributions. The radiation from internal pion lines is one such graph which we miss; this channel mainly contributes at large masses \cite{Shyam:2010vr}. Furthermore, the subthreshold $\rho$ production via resonances could be underestimated on the neutron by our model. Analogous to the $\eta$ case, it might be enhanced over $\rm pp\rightarrow\rho^0X$. And finally, channels like $\rm np\rightarrow d\rho^0$ (related to the radiative capture) could contribute, which are completely unknown.

The discrepancy of data and theory for the d+p reaction is specific for this reaction at this particular energy; the results for nuclear collisions to be discussed later do not show such a disagreement. We note that the observed cross section represents only about 15 - 20\% of the actual cross section; the rest is being cut away by the acceptance filter. Thus, any deficiencies, for example, in the angular distribution of our dileptons could show up in rather large errors of the spectra after the acceptance cuts have been performed.


\subsection{p + p at 3.5 GeV}

\begin{figure}[b]
  \includegraphics[width=0.5\textwidth]{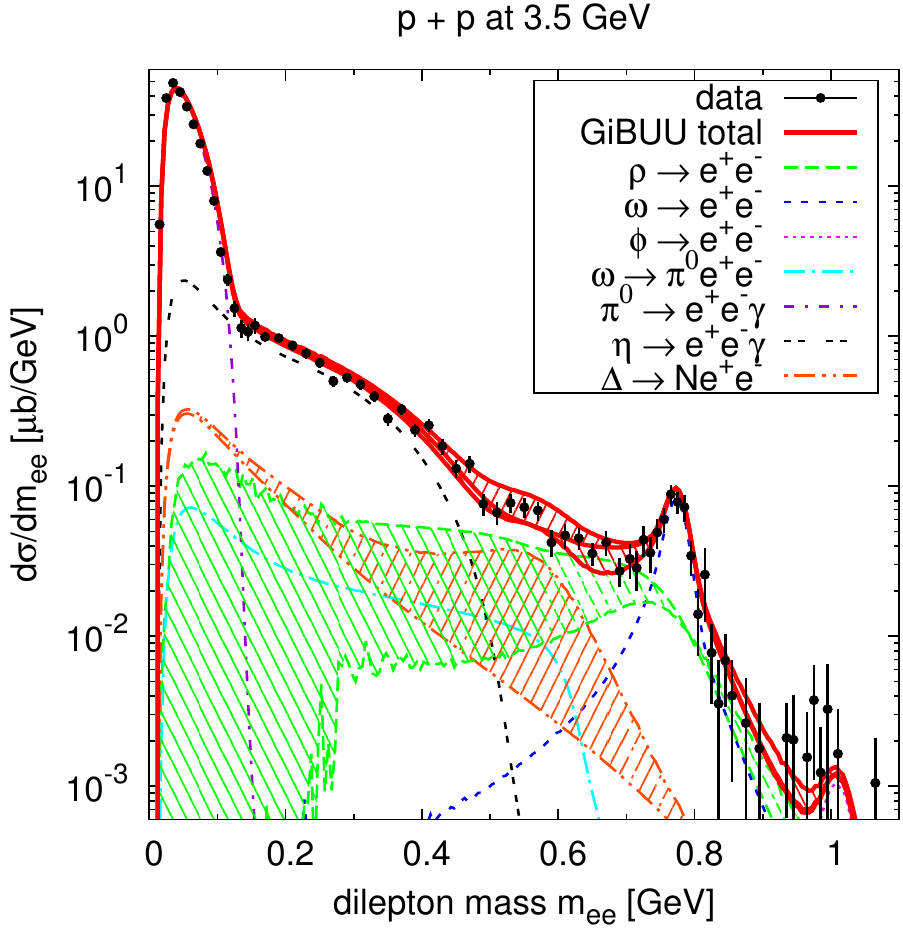}
  \caption{(Color online) Dilepton mass spectrum for pp collisions at 3.5\GeV. Data from \cite{HADES:2011ab}. The hatched areas indicate the effects of the $\Delta$ form factor \cite{Wan:2005ds} and baryon-resonance contributions to the  $\rho$ production, respectively. The total is shown (from bottom to top) with $\Delta$ form factor (left-hatched), $\rho$ resonance contributions (right-hatched) and with both of these effects together.}
  \label{fig:pp35_mass}
\end{figure}

Fig.~\ref{fig:pp35_mass} shows a comparison plot of a GiBUU simulation to HADES
data \cite{HADES:2011ab} for a proton beam of 3.5 \GeV
kinetic energy impinging on a fixed proton target. This is the highest beam energy (per nucleon) used by the HADES experiment and corresponds to a center-of-mass energy of $\sqrt{s}=3.18\GeV$.

\begin{figure}[t!]
  \includegraphics[width=0.5\textwidth]{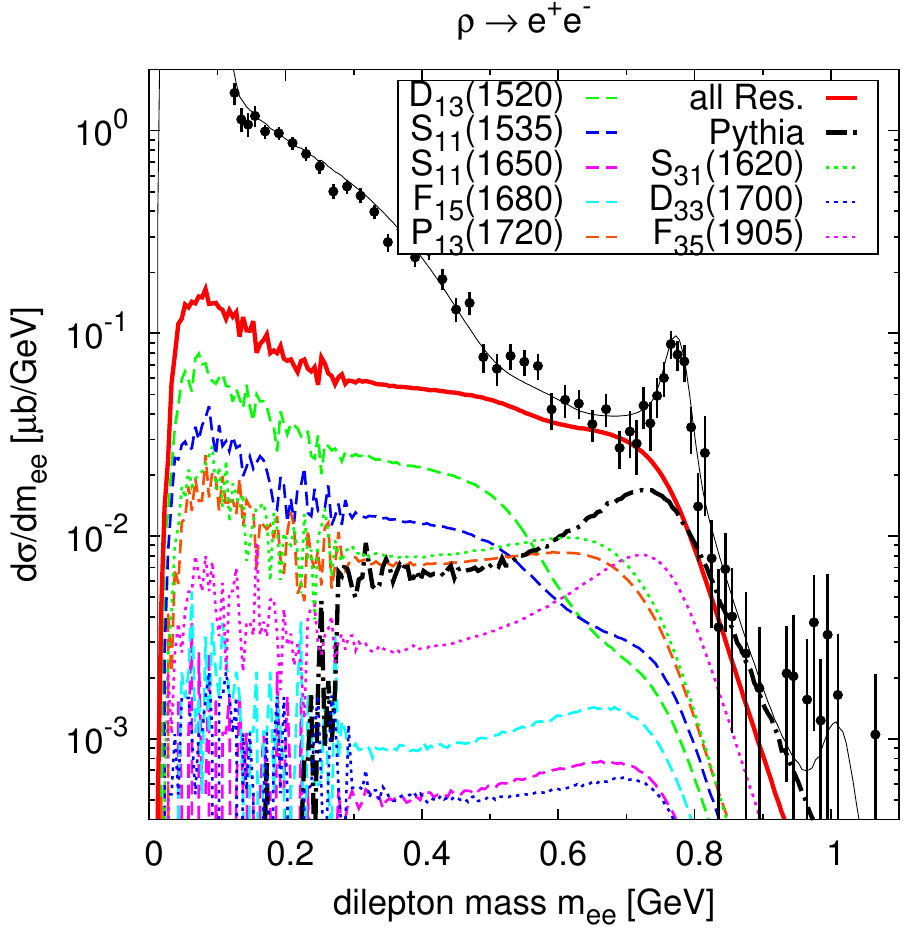}
  \includegraphics[width=0.5\textwidth]{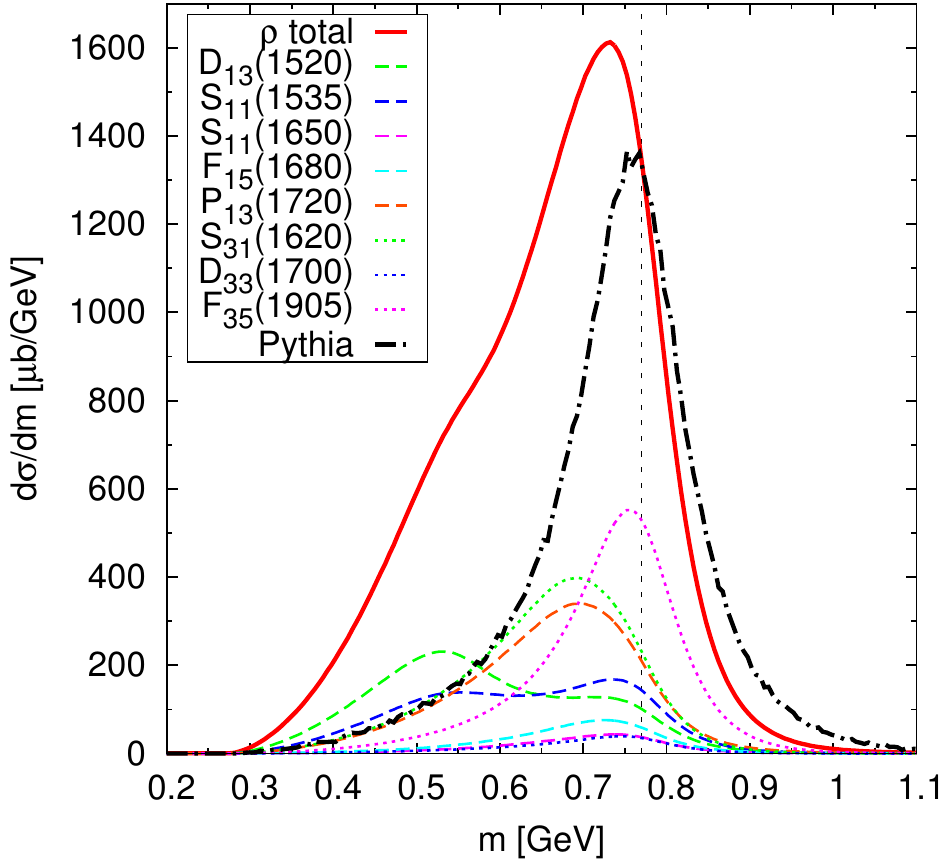}
  \caption{(Color online) Top: Resonance contributions to the $\rho$ channel in the dilepton mass spectrum. Bottom: Resonance contributions to the $\rho$ mass distribution. The dashed line indicates the vacuum pole mass of the $\rho$ meson. For comparison we also show the $\rho$ meson contribution from our earlier Pythia simulations \cite{Weil:2011fa}.}
  \label{fig:pp35_reso}
\end{figure}

At this energy, the $\eta$, $\omega$ and $\rho$ production channels are fully open, and even the $\phi$ production becomes energetically possible. The data only show a hint of a $\phi$ peak with very poor statistics, but it seems to be slightly underestimated by our simulation.

The $\eta$ and $\rho$ production is dominated by the channels $NN\rightarrow NN\pi\eta$ and $NN\rightarrow NN\pi\rho$, respectively. In our model these are saturated by double-resonance excitation, cf. sec. \ref{sec:RM}. The $\omega$ meson is presently produced in a non-resonant phase-space prescription through the exclusive and the $\pi\omega$ channel.

Under these assumptions, we get a very good agreement with the data over the whole mass range, as shown in fig.~\ref{fig:pp35_mass}.

\begin{figure}[b]
  \begin{center}
    \includegraphics[width=0.5\textwidth]{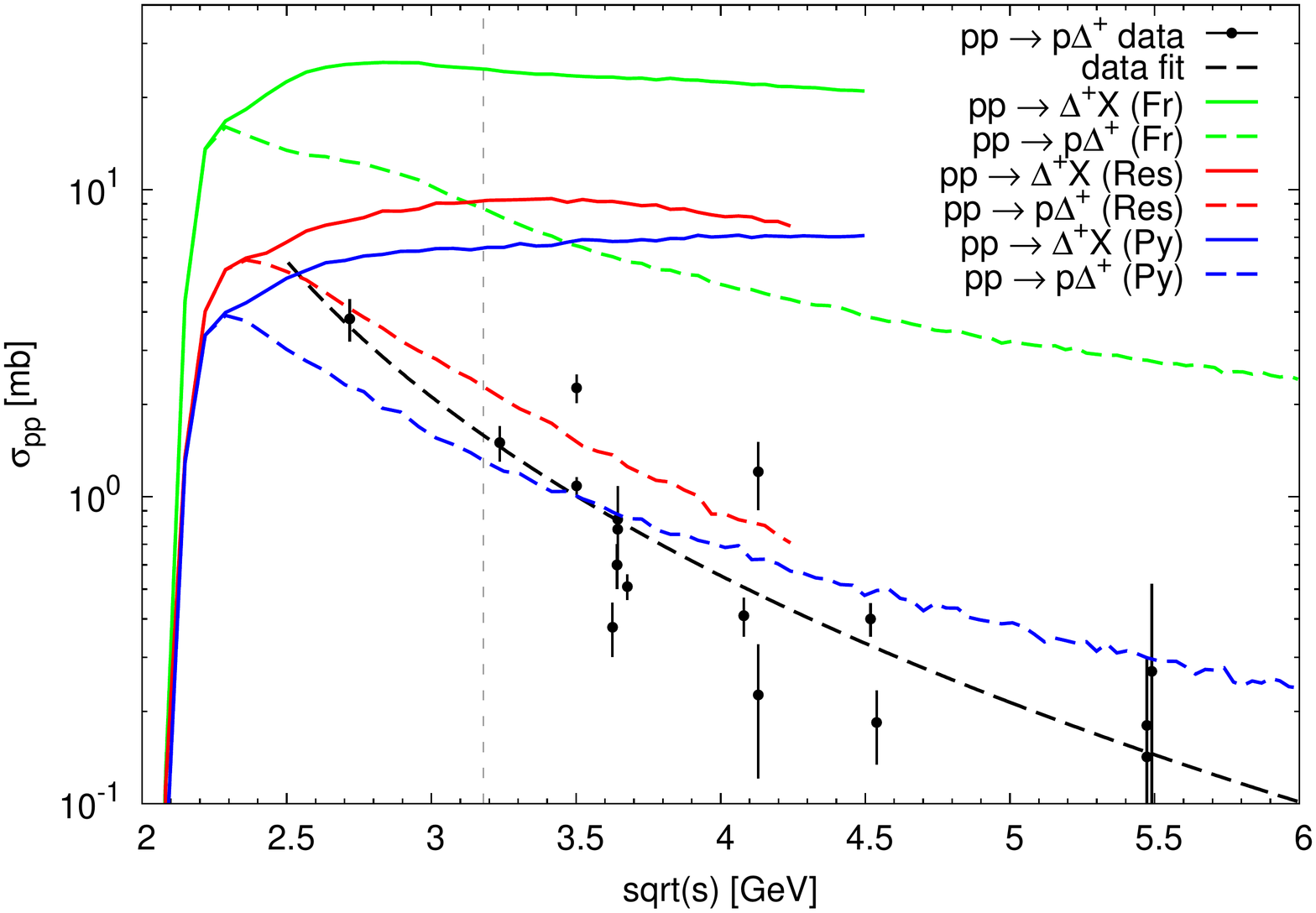}
  \end{center}
  \caption{(Color online) Inclusive and exclusive $\Delta^+$ production cross sections in different models (\Fritiof{} 7.02, \Pythia{} 6.4 and the GiBUU resonance model), compared to data from \cite{Baldini:1988ti}.}
  \label{fig:DeltaXS}
\end{figure}

It is interesting to note that the shape of the $\rho$ channel shown here differs significantly from the one obtained in our previous string-model investigations via \Pythia{} \cite{Weil:2011fa}, which has been adopted for the PLUTO simulations in \cite{HADES:2011ab}.
The latter is given by the lower dashed (green) line in Fig.\ \ref{fig:pp35_mass}, whereas the new resonance-model based treatment yields the upper dashed line. The $\rho$-shape effect is due to the production of $\rho$ mesons via nucleon resonances, i.e. $NN\rightarrow NR\rightarrow NN\rho$ and $NN\rightarrow \Delta R\rightarrow NN\pi\rho$, where the lighter resonances like e.g. $D_{13}(1520)$ will preferentially contribute to the low-mass part of the $\rho$ spectral function. Together with the $1/m^3$ factor of the dilepton decay width, this results in a very flat distribution, which lacks a clear peak at the nominal mass, and dominates the dilepton spectrum in the intermediate mass region around 500 - 700 \MeV.

\begin{figure*}[ht]
    \includegraphics[width=\textwidth]{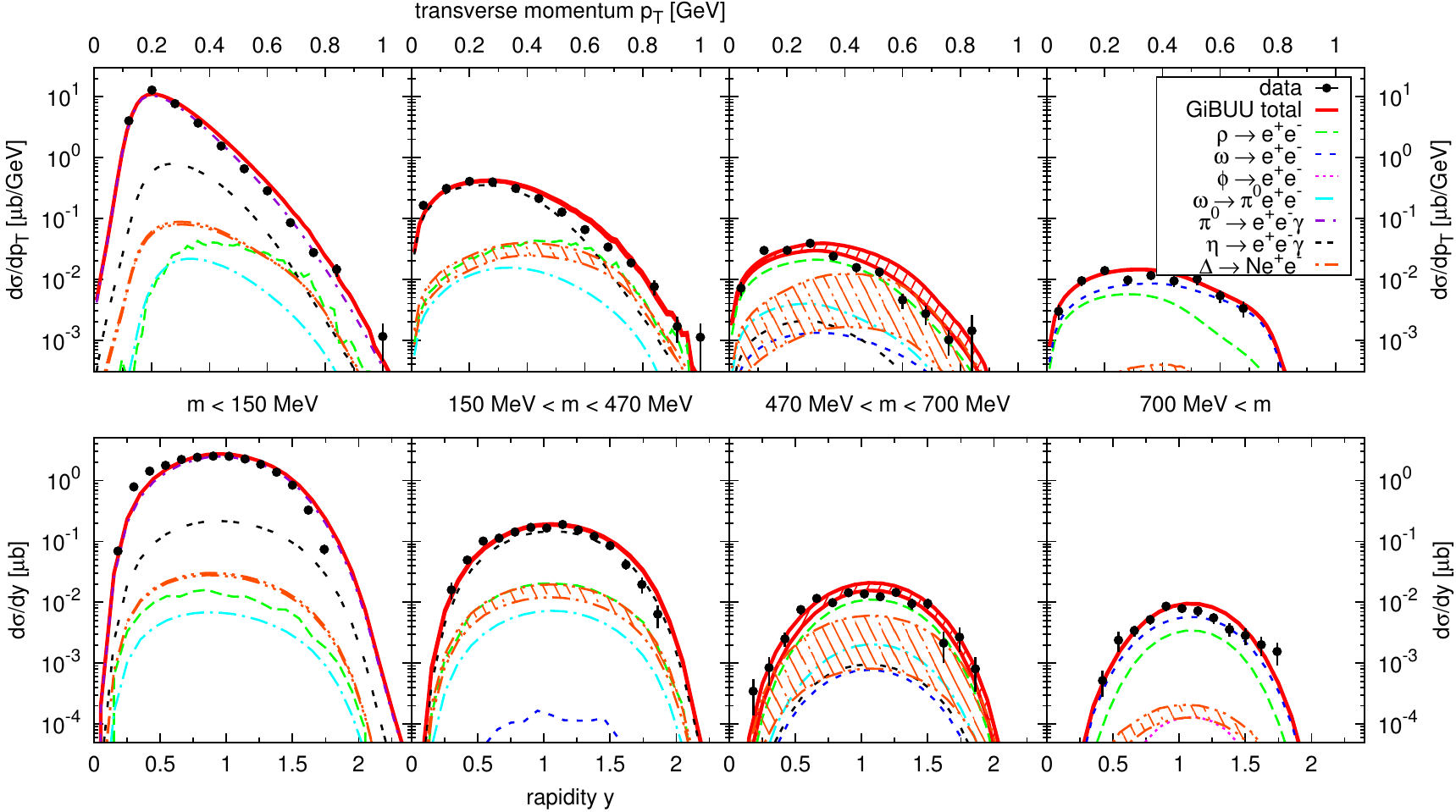}
  \caption{(Color online) Transverse momentum and rapidity spectra of dilepton pairs from pp at 3.5\GeV in four mass bins. The hatched area indicates the effect of the $\Delta$ form factor. Data from \cite{HADES:2011ab}.}
  \label{fig:pp35_pty}
\end{figure*}

The $\rho$ spectral function is thus `modified' already in the vacuum, simply due to the production mechanism via nucleon resonances. As seen in fig.~\ref{fig:pp35_reso}, the $\rho$ mass distribution in pp at 3.5\GeV peaks around 730\MeV, with an additional shoulder around 500\MeV (due to low-mass resonances, mainly the $D_{13}(1520)$). This spectral shape is due to phase-space limitations and special resonance properties. It differs significantly from the mass distribution resulting from a \Pythia{} simulation \cite{Weil:2011fa}, which lacks any resonance contributions. Similar effects were already observed, e.g., in C+C reactions \cite{Schumacher:2006wc}. We stress here that this is not an `in-medium' effect at all: It is solely caused by the production mechanism and occurs already in elementary p+p collisions in the vacuum. This effect is crucial for understanding the intermediate mass region of the dilepton spectrum in pp collisions at 3.5\GeV (as seen in fig.~\ref{fig:pp35_mass}) and might also play an 
important role at 2.2\GeV (see next section).

The particular influence of the $N^*(1520)$ resonance on dilepton spectra from NN collisions have already been investigated in \cite{Bratkovskaya:1999mr}, where it was concluded that the $N^*(1520)$ can indeed give sizable contributions to the DLS and HADES spectra, but is subject to moderate uncertainties.

It should be noted that the exact composition of the resonance contributions to the $\rho$ channel, and therefore also its exact shape, are not fixed by data so far, but rather represent an `educated guess'. The resonance composition can be checked via $\pi N$ invariant mass spectra.

Moreover, possible $\rho\Delta$ decay modes of certain resonances could give further contributions to the dilepton cocktail, as mentioned earlier.

Comparing our cocktail to other transport models like HSD \cite{Bratkovskaya:2007jk} or UrQMD \cite{Schmidt:2008hm}, one of the most significant discrepancies shows up in the size of the $\Delta$ channel. While in our model the $\Delta$ does not give any significant contribution to the total dilepton yield at $E_{\rm kin}=3.5\GeV$ (without a form factor), this is not so for the two other models. Both of them have a much stronger $\Delta$ channel, which even dominates the dilepton spectrum in the intermediate mass region around 600\MeV. We stress here that there are several factors of uncertainty in the $\Delta$ channel, for example the inclusive production cross section, but also the parametrization of the $\Delta$ decay width (hadronic as well as leptonic) and the completely unsettled question of the electromagnetic N-$\Delta$ transition form factor.

Although the inclusive $\Delta$ production cross section is not that well known at $E_{\rm kin}=3.5\GeV$, one can get constraints from the exclusive cross section, cf.~fig.~\ref{fig:DeltaXS}, as well as the inclusive one at lower energies (where it is fixed via pion production). Both constraints are respected in our resonance model, while e.g. the \Fritiof{} model clearly overestimates the exclusive $\Delta^+$ production, and in particular does not seem to respect the correct isospin relations.

On the question of the electromagnetic N-$\Delta$ transition form factor, it should be noted that in our simulations the Iachello model \cite{Wan:2005ds} agrees reasonably well with the data (depending on the contributions of other baryonic resonances), while a naive VMD form factor, as used e.g.~in \cite{Titov:1994vg}, would clearly overshoot the data.

In order to understand the underlying processes, it is not sufficient
to consider only the mass spectrum. Other observables can give further
insight into the reaction dynamics and can serve as a cross check for the
validation of theoretical models. In order to compare to the data from \cite{HADES:2011ab},
we examine the transverse momentum
and rapidity distributions in four different mass bins
(see fig.~\ref{fig:pp35_pty}):

\begin{figure}[b]
  \includegraphics[width=0.5\textwidth]{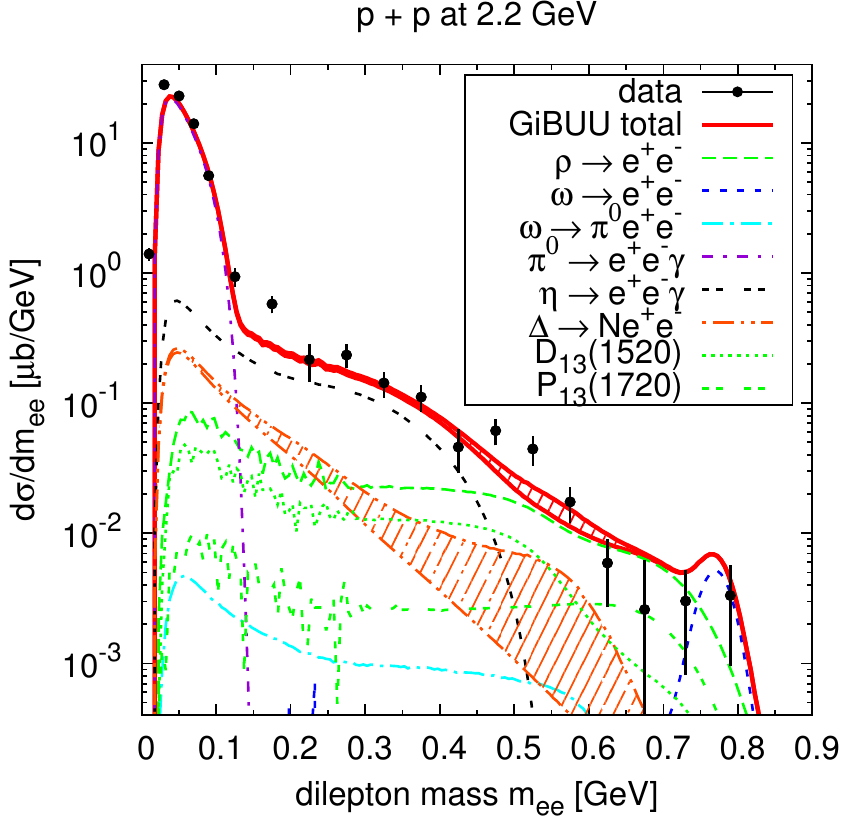}
  \caption{(Color online) Dilepton mass spectrum for pp at 2.2\GeV. The hatched area indicates the effect of the $\Delta$ form factor. Data from \cite{Agakishiev:2012tc}.}
  \label{fig:pp22_mass}
\end{figure}

\begin{figure*}[ht]
  \includegraphics[width=\textwidth]{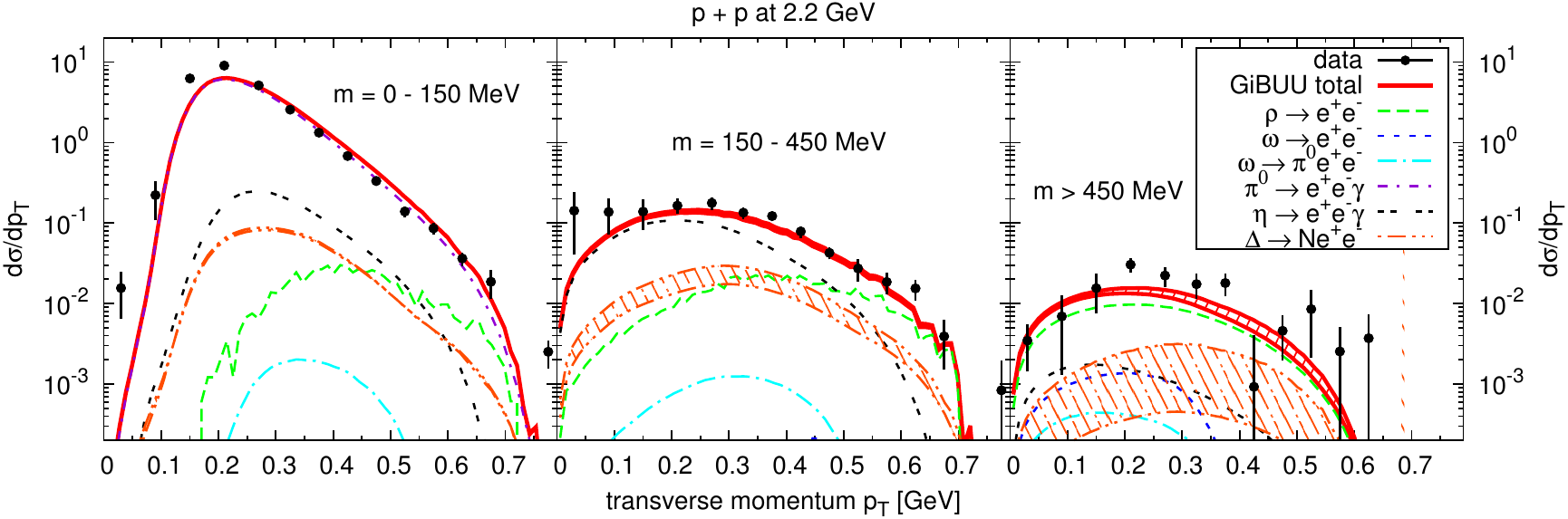}
  \caption{(Color online) Transverse momentum spectra of dilepton pairs from pp at 2.2\GeV in three mass bins. The hatched area indicates the effect of the $\Delta$ form factor. Data from \cite{Agakishiev:2012tc}.}
  \label{fig:pp22_pt}
\end{figure*}

\begin{itemize}
\item $m < 150\MeV$, dominated by the $\pi^0$ Dalitz channel,
\item $150\MeV<m<470\MeV$, dominated by the $\eta$ Dalitz decay,
\item $470\MeV<m<700\MeV$, dominated by the direct $\rho$ decay (possibly with contributions from the $\Delta$ Dalitz),
\item $700\MeV<m$, dominated by the $\omega$ and $\rho$.
\end{itemize}

Distinguishing several mass bins is useful in order to separate the
contributions of different channels.
In all four mass bins, we achieve an excellent agreement with
the HADES data \cite{HADES:2011ab}. In particular it should be noted that a stronger $\Delta$ channel would apparently destroy the very good  agreement in the $p_T$ spectra, since it would yield too large high-$p_T$ contributions in the two mass bins of 150-470 and 470-700\MeV.


\subsection{p + p at 2.2 GeV}

A third, intermediate, beam energy of 2.2\GeV has been used for the HADES experiment. This energy is well above the $\eta$ production threshold and is just high enough to reach the pole mass of the light vector mesons, $\rho$ and $\omega$, which dominate the high-mass part of the dilepton spectrum (as seen in fig.~\ref{fig:pp22_mass}).
The $\Delta$ channel plays a less important role here, since it is buried underneath the strong $\eta$ and $\rho$ channels. The $\omega$ only gives a small contribution, since the energy is only just at the threshold of $\omega$ production.

The $\rho$ channel exhibits slightly more structure here than at 3.5 \GeV, showing a moderate step around 550 \MeV. This step marks the border between a low-mass part, which is dominated by the $D_{13}(1520)$ resonance, and a high-mass part dominated by the $P_{13}(1720)$. In fig.~\ref{fig:pp22_mass} we show the contributions of these two resonances to the $\rho$ channel, but omit the subdominant contributions of other resonances (for the sake of readability). The resonance contributions indeed improve the agreement with the data, compared to the PLUTO cocktail, which only includes phase-space population of the $\rho$ \cite{Agakishiev:2012tc}. However, there are still minor deviations, which seem to suggest an underestimation of the $D_{13}(1520)$ and an overestimation of the $P_{13}(1720)$ in our resonance cocktail at this energy.

In fig.~\ref{fig:pp22_pt} we show the $p_T$ spectra for three different mass bins in comparison to the data from \cite{Agakishiev:2012tc}. Our simulations give a better agreement with the data than the PLUTO cocktail shown in \cite{Agakishiev:2012tc} in all three mass bins. Most notably, we get an improvement from the larger $\rho$ contribution in the highest mass bin.


\subsection{Comparison to elementary DLS data}

\begin{figure*}[p]
  \includegraphics[width=\textwidth]{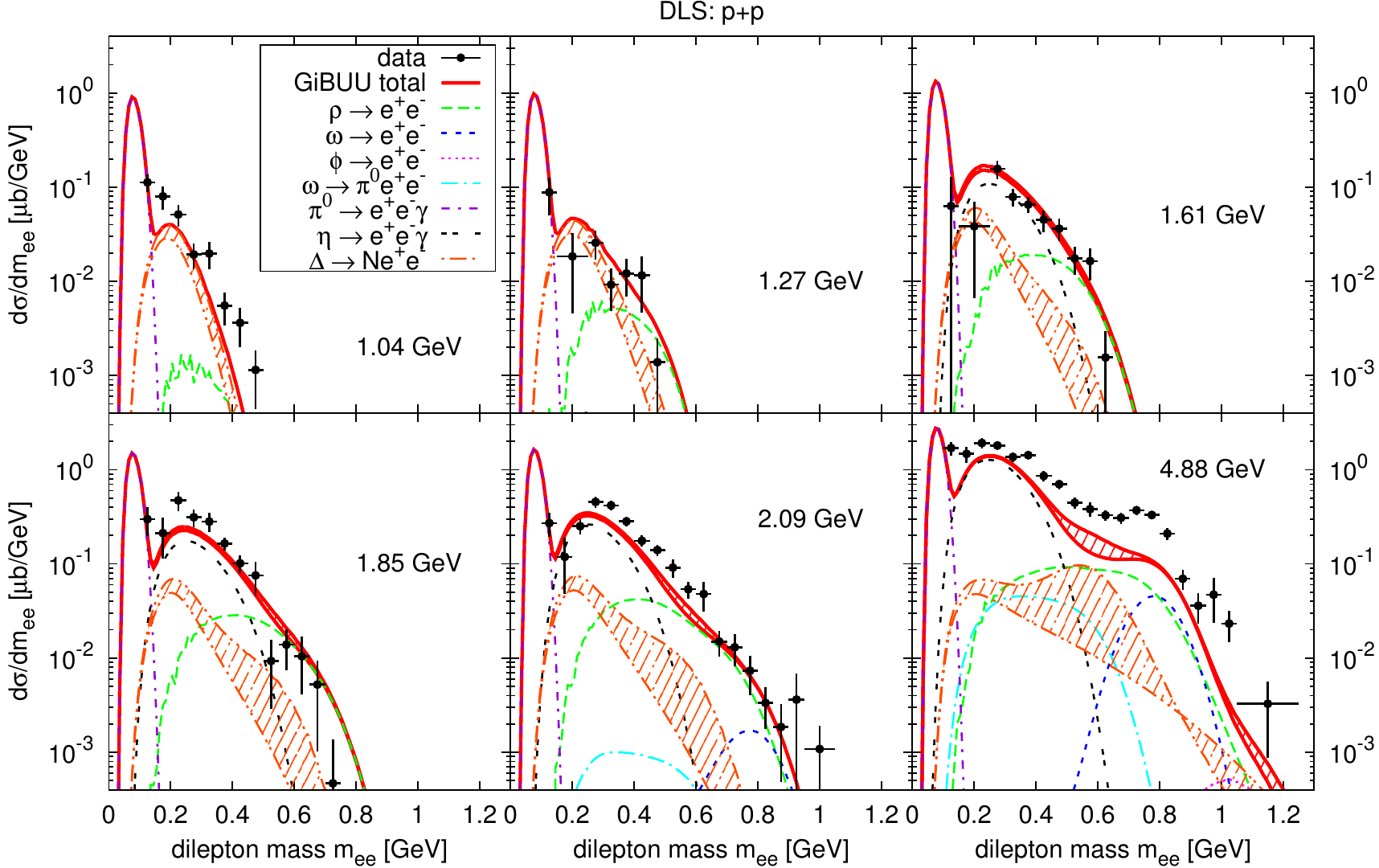}
  \includegraphics[width=\textwidth]{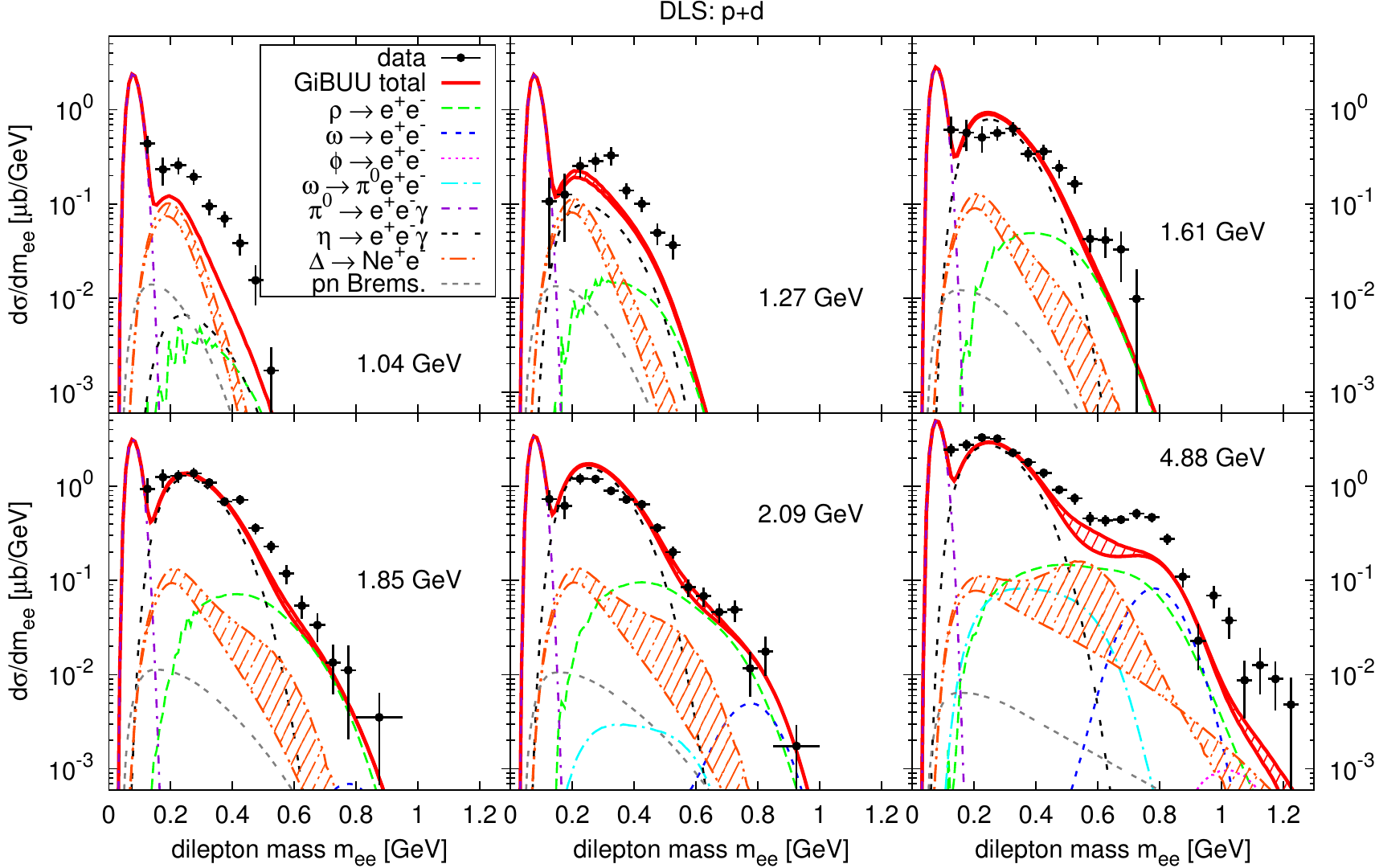}
  \caption{(Color online) Dilepton mass spectra in comparison to DLS data \cite{Wilson:1997sr}. Top: p+p, bottom: p+d.}
  \label{fig:DLS_mass}
\end{figure*}

In addition to the recently measured HADES data, also the elementary data measured previously by the DLS collaboration are available for comparison with our model \cite{Wilson:1997sr}. Unfortunately they are of inferior quality in terms of statistics and acceptance. However, more beam energies have been measured than in the case of HADES, so that they can still provide additional contraints, which are useful for understanding the elementary cocktail.

In order to compare to the DLS data, the GiBUU dilepton events have been filtered throug the DLS acceptance filter, version 4.1, as available from \cite{dls}. In addition to the acceptance filtering, the events have been smeared with a Gaussian of width $\sigma=0.1m_{\rm ee}$, in order to account for the mass resolution of the detector. No further cuts have been applied. The kinematics of the reactions measured by DLS are summarized in table \ref{tab:DLS_reactions}. At each of the given energies, a p+p and p+d reaction was measured.

\begin{table}[h]
  \begin{center}
    \begin{tabular}{|c|c|c|}
      \hline
      $E_{\rm kin}$ & $\sqrt{s}$ & $p_{\rm lab}$  \\
      \hline
      1.04 & 2.34 & 1.74 \\
      1.27 & 2.43 & 2.00 \\
      1.61 & 2.56 & 2.37 \\
      1.85 & 2.64 & 2.63 \\
      2.09 & 2.73 & 2.88 \\
      4.88 & 3.56 & 5.74 \\
      \hline
    \end{tabular}
  \end{center}
  \caption{Kinematic conditions of the elementary collisions measured by DLS (in GeV).}
  \label{tab:DLS_reactions}
\end{table}

The comparison of the GiBUU model results to the DLS data is shown in Fig.~\ref{fig:DLS_mass}. As before, we show the effect of the $\Delta$ transition form factor as a hatched band and note that it slightly improves the agreement with the data in almost all cases.

Is is apparent that at the medium beam energies there is a reasonable agreement, both in p+p and p+d. The largest deviations are visible at the highest beam energy of 4.88\GeV, which is already at the border of validity of our resonance model. Apparently the inclusive production of $\rho$ and $\omega$ mesons is underestimated there.

The underestimation at the lowest energy of 1.04\GeV is similar to that seen in the HADES experiment at a comparable energy (see Fig.~\ref{fig:dp125}). Since at this somewhat lower energy the $\eta$ production plays no role, the discrepancy seems to indicate a problem with the $\Delta$ or Bremsstrahlung contributions. However, we note again that the population of the $\Delta$ resonance is constrained rather well by the pion and total cross sections, which we describe rather well (see Fig.~\ref{fig:NN_inel}). On the other hand, the decay of the resonance is fixed by the electromagnetic coupling at the photon point, so that there is no ambiguity there. Further, form factors have only little influence at such low eneries. We thus have to conclude that we have no explanation for the discrepancy yet and note that related, earlier calculations similarly underestimated the DLS dilepton yield at this lowest energy \cite{Shekhter:2003xd}. 


\section{Dilepton spectra from p+Nb collisions}

\begin{figure}[t]
  \includegraphics[width=0.5\textwidth]{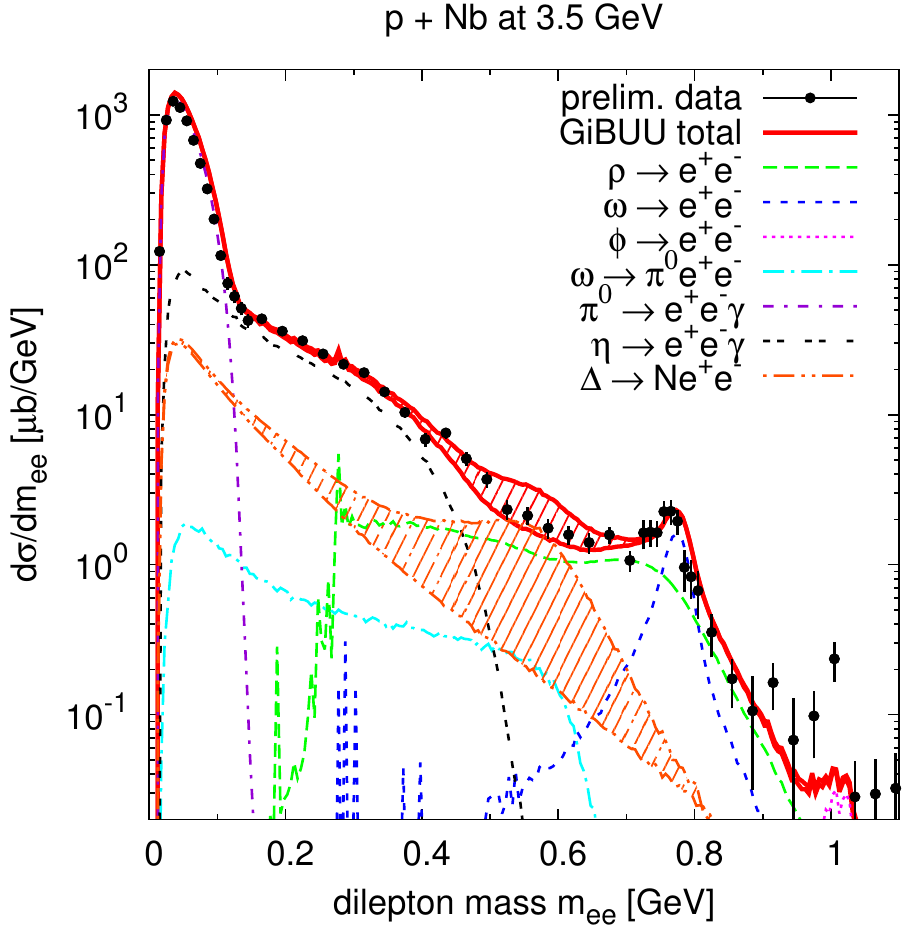}
  \caption{(Color online) Dilepton mass spectrum for p+Nb at 3.5\GeV,
    showing all contributing channels with vacuum spectral functions. The hatched area indicates the effect of the $\Delta$ form factor. Preliminary data taken from \cite{Weber:2011zz}, scaled to fit the $\pi^0$ and $\eta$ yield.}
  \label{fig:pNb_mass_1}
\end{figure}

Fig.~\ref{fig:pNb_mass_1} shows simulated dilepton spectra for p+Nb collisions at
3.5 \GeV using vacuum spectral functions, compared to preliminary data from \cite{Weber:2011zz}. As for p+p at 3.5 \GeV, we filter our dilepton events through the HADES acceptance filter and cut on $0.08\GeV<p_{\rm lep}<2.0\GeV$ and $\theta_{\rm ee}>9\degree$.
The level of agreement is similar to the p+p reaction at the same energy. Note, however, that the data are not absolutely normalized in terms of a cross section yet. Therefore we have scaled the data points to match the simulation in the low-mass region, which is dominated by the $\pi^0$ and $\eta$ Dalitz channels. Moreover, the data have not been fully corrected for all detector effects yet, which is the reason for a slight shift of the mass scale (on the order of 1\%), which is visible at the $\omega$ peak \cite{lorenz_priv}. It is evident that the data can be quite well described if the electromagnetic $\Delta$ decay width does not contain the form factor of ref.~\cite{Wan:2005ds} which would create a hump in the spectrum around 0.6 GeV.

In contrast to the NN collisions in the preceding chapter, here we neglect the $\rho$-meson contributions below the $2\pi$ threshold (due to numerical reasons). As seen in fig.~\ref{fig:pNb_mass_1}, they do not contribute significantly to the total dilepton yield.

\subsection{In-medium effects}

\begin{figure*}[t]
  \includegraphics[width=\textwidth]{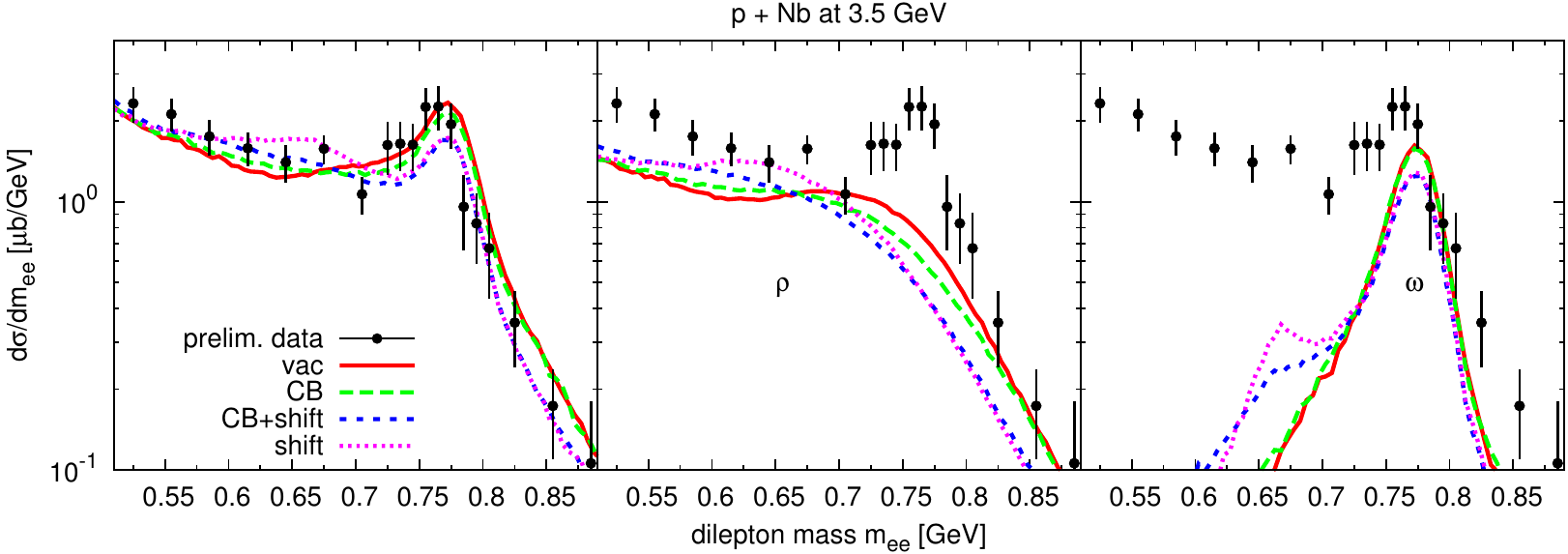}
  \caption{(Color online) Dilepton mass spectra for p+Nb at 3.5\GeV.
    Comparison of different in-medium scenarios (vacuum spectral functions for the vector mesons,
    collisional broadening, $16\%$ mass shift, collisional broadening plus mass shift).
    Left: Total spectrum, center: $\rho$ contribution, right: $\omega$ contribution.
    Preliminary data taken from \cite{Weber:2011zz}, scaled to fit the $\pi^0$ and $\eta$ yield.}
  \label{fig:pNb_mass_2}
\end{figure*}

In p+Nb reactions there are additional effects,
compared to the elementary p+p reactions. First of all, the primary p+N
collisions will be nearly identical, apart from binding effects and
some Fermi smearing and Pauli blocking, but besides p+p also p+n collisions play a
role.  Furthermore, the produced particles undergo final-state
interactions within the Nb nucleus, and processes like meson absorption
and regeneration may become important. The secondary collisions will
on average have lower energies than the primary N+N collisions. Finally also the vector-meson spectral
functions may be modified in the nuclear medium.

The propagation of particles with density-de\-pendent spectral functions
(usually referred to as ``off-shell propagation'') poses a particular challenge.
Our approach to this problem is based on the off-shell equations of motion of
test particles, as given in \cite{Cassing:1999wx} and \cite{Leupold:1999ga}. Such an off-shell treatment
is necessary for including in-medium modifications of the spectral functions
(e.g. collisional broadening of the vector mesons). The collisional width inside
a nuclear medium of density, $\rho$, can be related to the collision cross section,
$\sigma_{NX}$, in low-density approximation as

\begin{equation}
 \Gamma_{\rm coll} = \rho \left< v_{\rm rel} \sigma_{NX} \right> \; ,
\end{equation}

where $v_{\rm rel}$ is the relative velocity and the brackets indicate an
integration over the Fermi momentum of the nucleons. This collisional
width will in general depend on the momentum of the involved particle, $X$. In order to avoid numerical difficulties connected with the appearance of superluminous test particles, we neglect the momentum
dependence and use the simplified form,

\begin{equation}
 \Gamma_{\rm coll} = \Gamma_0 \frac{\rho}{\rho_0} \; ,
\end{equation}

where $\rho_0=0.168\fm^{-3}$ is the normal nuclear matter density. The value of
$\Gamma_0$ should on average match the mo\-men\-tum-dependent width as obtained from
the collision term. We typically use $\Gamma_0 = 150 \MeV$ for the $\rho$ and
$\Gamma_0 = 80 \MeV$ for the $\omega$ meson. More details on off-shell propagation
in the GiBUU model in general can be found in \cite{Buss:2011mx}.

The mass spectrum above 500 \MeV can receive modifications from
the inclusion of in-medium effects in the vector-meson spectral
functions. Fig.~\ref{fig:pNb_mass_2} shows the typical in-medium scenarios:
The first one includes a collisionally broadened in-medium width, while the second one assumes a pole-mass shift
according to
\begin{equation}
 m^*(\rho) = m_0\left(1-\alpha\frac{\rho}{\rho_0}\right) \; ,
\end{equation}
with a scaling parameter, $\alpha = 16 \%$.
The third scenario combines both of these effects. The modifications
introduced by these scenarios are roughly on the same order of magnitude
as the systematic errors of the data, and so far there is no clear evidence for medium modifications of the vector-meson properties in cold nuclear matter from the HADES data. However, it looks as if a mass shift tends to deteriorate the agreement with the data.

Regarding the $\omega$ absorption, it should be noted that the GiBUU implementation yields an average collisional width of roughly $\Gamma_0=80\MeV$. This appeared too low to explain the transparency-ratio measurement of \cite{:2008xy}, which seemed to demand values of 130 to 150 \MeV. For the HADES dilepton data, such a discrepancy currently does not seem to exist.

However, one should keep in mind that a statement about $\omega$ absorption depends on a number of prerequisites. For example, one needs to have the $\rho$ contribution well under control, since it represents a large background under the $\omega$ peak. Given the discussion about resonance contributions to the elementary $\rho$ production, this is already not a trivial task, even more complicated by possible in-medium modifications of the $\rho$ meson. Furthermore, the size of the $\omega$ peak in pNb crucially depends not only on the production cross section in pp collisions (which is well determined via the elementary pp data at 3.5\GeV), but also in pn, which is unknown. We assume $\omega$ production cross sections which are isospin-independent, i.e. equal in pp and pn.

\begin{figure*}[hp]
  \begin{center}
   \includegraphics[width=\textwidth]{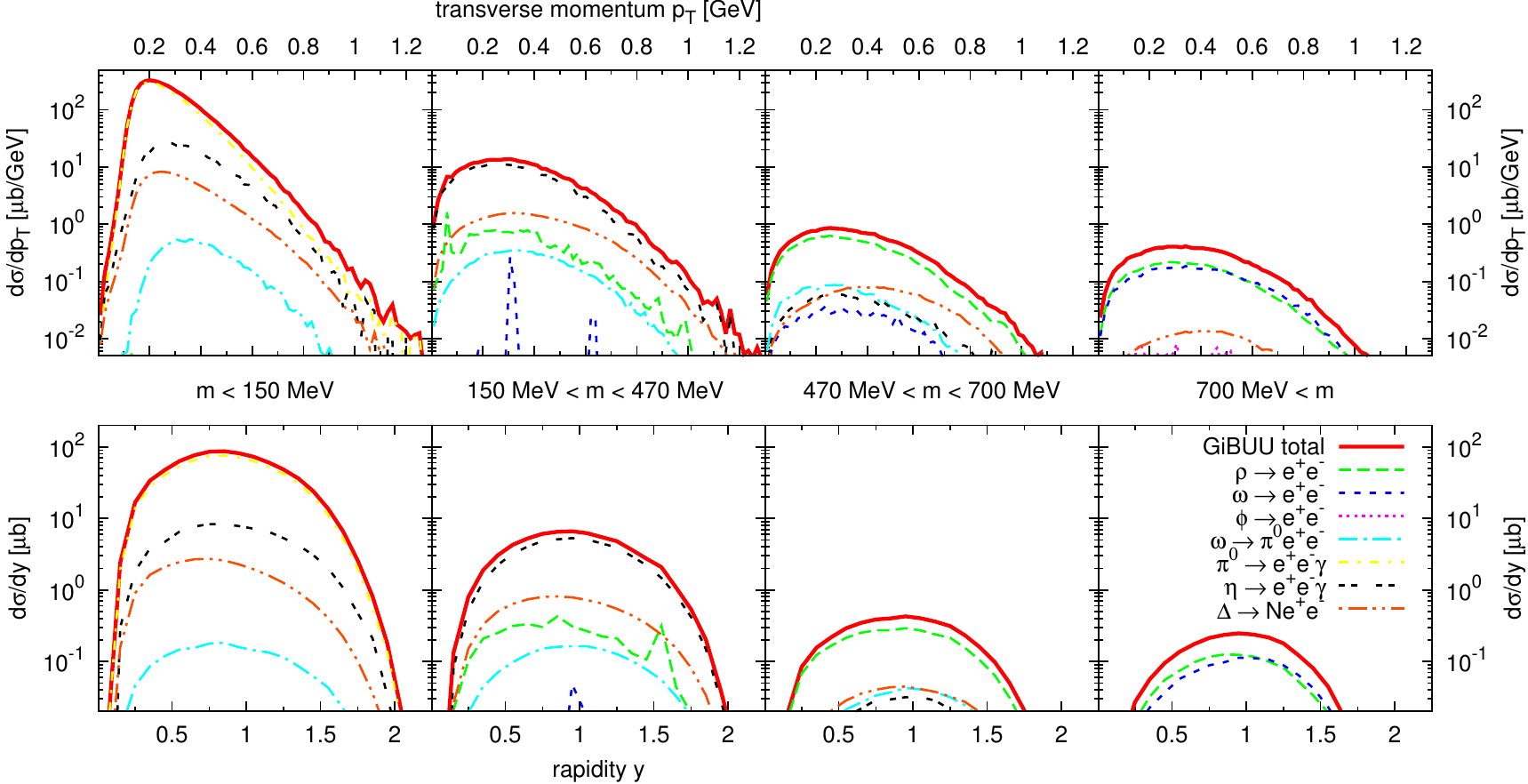}
  \end{center}
  \caption{(Color online) $p_T$ and rapidity spectra of dileptons from p+Nb reactions in four mass bins.}
  \label{fig:pNb_pt_y}
\end{figure*}

\begin{figure*}[hp]
  \centering
  \includegraphics[width=0.9\textwidth]{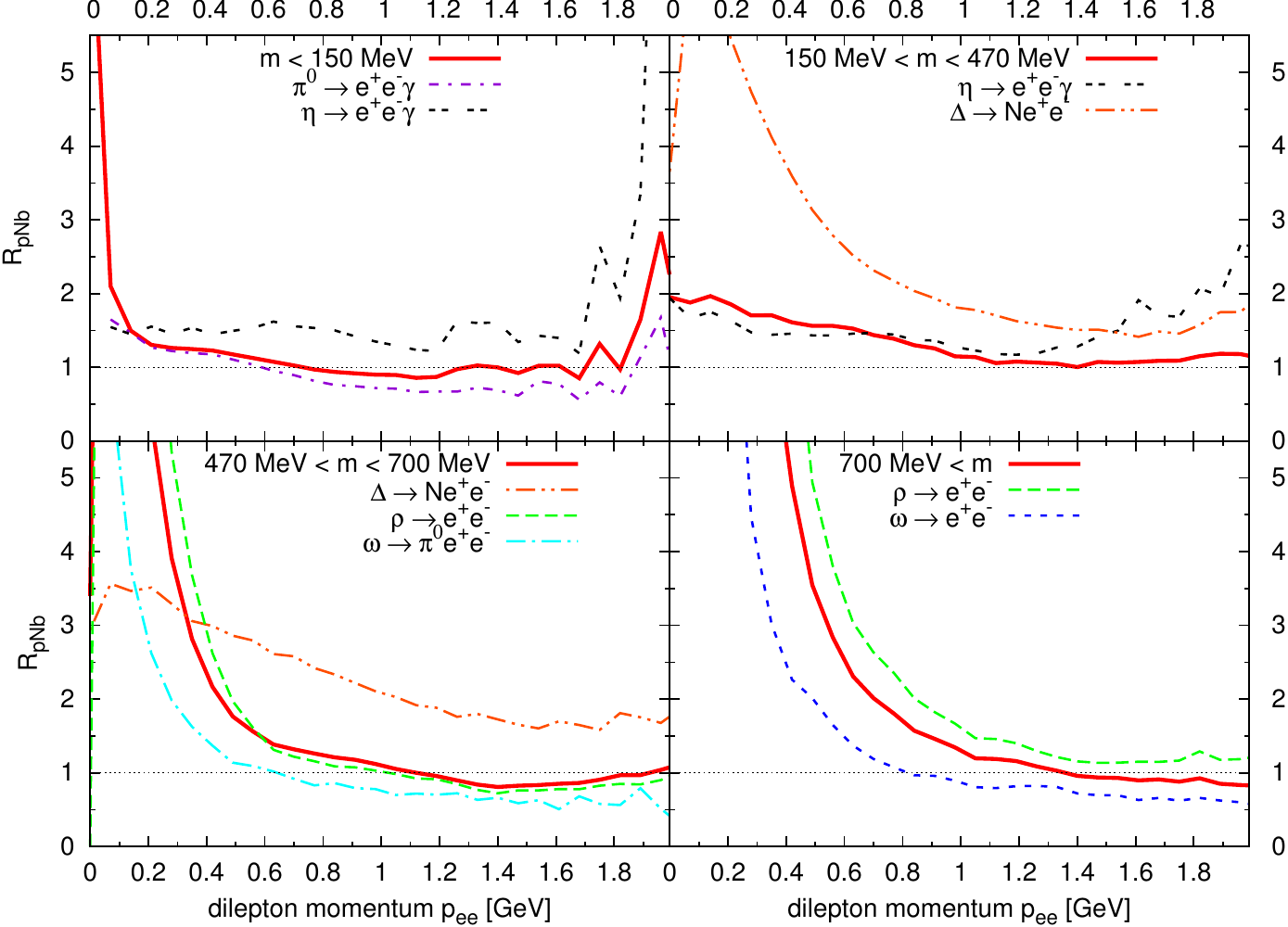}
  \caption{(Color online) Ratio of dilepton yield from p+Nb and p+p collisions at 3.5\GeV as a function of momentum, in four mass bins.}
  \label{fig:R_pNb}
\end{figure*}

In addition to the in-medium modifications of the vector mesons, also the baryonic resonances can receive similar modifications in the medium. Since the production via baryon resonances is particularly important for the $\rho$ meson, in-medium modifications of these resonances can lead to further modifications of the $\rho$ contribution to the dilepton spectrum, which should be considered in future investigations.

The $p_T$ and rapidity spectra for p+Nb are depicted in fig.~\ref{fig:pNb_pt_y}
with the same mass binning as in the p+p case.
The shown $p_T$ and rapidity spectra do not include any in-medium effects for
the vector mesons and are not significantly sensitive to such modifications.

For a further discussion of the nuclear effects, it is useful to consider the quantity

\begin{equation}
 R_{\rm pNb} = \frac{\sigma_{\rm pNb\rightarrow e^+e^-X}}{\sigma_{\rm pp\rightarrow e^+e^-X}} \cdot \frac{\sigma_{\rm pp\rightarrow X}}{\sigma_{\rm pNb\rightarrow X}} \; ,
\end{equation}

i.e., the ratio of dilepton yields in pNb vs.~pp, normalized to the total cross section for these reactions (whose ratio is roughly $\sigma_{\rm pNb\rightarrow X}/\sigma_{\rm pp\rightarrow X}\approx25.0$ in our simulations). If medium effects are negligible, this quantity will be unity. Therefore, any deviation from unity indicates medium effects such as, e.g., absorption ($R<1$) or secondary production ($R>1$). Fig.~\ref{fig:R_pNb} shows $R_{\rm pNb}$ as a function of the dilepton momentum in four different invariant-mass bins, with the contributions from the different source channels.

While $R_{\rm pNb}$ is relatively flat in the $\pi^0$ region, the higher mass bins show a strong enhancement at low momenta, which can be understood as secondary particle production and/or elastic rescattering. The high momentum region in all mass bins tends to show a slight depletion, connected to absorption.

The observable $R_{\rm pNb}$ could also help to pin down the relative contribution of the $\Delta$ Dalitz channel to the dilepton spectrum. As can be seen in Fig.~\ref{fig:R_pNb}, the $\Delta$'s ratio is rather large, due to the enhanced production of the $\Delta^{+,0}$ charge states in pn collisions, relative to pp. The isospin factors for $NN\rightarrow N\Delta^{+,0}$ are a factor of two larger in pn than in pp.

This isospin dependence could provide additional constraints for distinguishing the $\rho$ and $\Delta$ contributions in the intermediate mass range of 470 - 700 MeV. Since the $\rho$ channel dominates our simulated cocktail in this mass range (without a $\Delta$ form factor), the total value of $R_{\rm pNb}$ roughly follows the $R$-value of the $\rho$ channel. If the spectrum would be dominated by the $\Delta$ Dalitz channel in this mass range, then the total value of $R_{\rm pNb}$ would be more similar to the $\Delta$'s $R$-value.


\section{Conclusions}

We have shown that the HADES data from elementary N+N collisions can be described consistently by an extended resonance model over the whole range of beam energies. We have set up such a model based on the earlier resonance model approach by Teis et al.

For describing the dilepton mass spectrum at the highest beam energy of 3.5\GeV, an essential ingredient is a $\rho$ spectral function, which is modified through the production via nucleon resonances, with an enhanced low-mass contribution from low-lying resonances like the $D_{13}(1520)$.

After fixing the model with the constraints given by the elementary N+N collisions, the p+Nb reaction at 3.5\GeV is reasonably well described by the GiBUU transport model, using the same input and without requring any in-medium mass shifts. According to our model, the p+Nb data show only a limited sensitivity to collisional broadening of the $\rho$ meson.

These results also provide the basis for a further investigation of the heavy-ion collisions at SIS energies measured by the HADES collaboration \cite{Agakichiev:2006tg,Agakishiev:2007ts,Agakishiev:2011vf}.


\section*{Acknowledgments}

We thank the HADES collaboration for providing us with the data and the HADES
acceptance filter and for many fruitful discussions. Special thanks go to
Tetyana Galatyuk, Anar Rustamov, Manuel Lorenz and Malgorzata Gumberidze. Moreover we are grateful to Kai Gallmeister and Volker Metag for many stimulating discussions of the topics presented here.
This work was supported by HIC4FAIR, HGS-HIRe and BMBF.


\bibliographystyle{epj}
\bibliography{references}

\end{document}